\documentclass{amsart}
\usepackage{amssymb}
\usepackage{amscd}

\begin{document}
\title{ENTANGLEMENTS AND COMPOUND STATES IN QUANTUM INFORMATION THEORY}
\author{Viacheslav P Belavkin}
\address{Department of Mathematics, University of Nottingham, NG7 2RD Nottingham, UK}
\thanks{The first author acknowledge JSPS Visiting Fellowship support and UK--Japan
Royal Society joint research grant and hospitality of Science University of Tokyo.}
\author{Masanori Ohya}
\address{Department of Information Sciences, Science University of Tokyo, 278 Noda
City, Chiba, Japan}
\date{July 20, 1999}
\maketitle

\begin{abstract}
Quantum entanglements, describing truly quantum couplings, are studied and
classified from the point of view of quantum compound states. We show that
classical-quantum correspondences such as quantum encodings can be treated
as d-entanglements leading to a special class of the separable compound
states. The mutual information of the d-compound and entangled states lead
to two different types of entropies for a given quantum state: the von
Neumann entropy, which is achieved as the supremum of the information over
all d-entanglements, and the dimensional entropy, which is achieved at the
standard entanglement, the true quantum entanglement, coinciding with a
d-entanglement only in the commutative case. The q-capacity of a quantum
noiseless channel, defined as the supremum over all entanglements, is given
as the logarithm of the dimensionality of the input von Neumann algebra. It
can double the classical capacity, achieved as the supremum over all
semi-quantum couplings (d-entanglements, or encodings), which is bounded by
the logarithm of the dimensionality of a maximal Abelian subalgebra.
\end{abstract}

\section{Introduction}

Recently, the specifically quantum correlations, called in quantum physics
entanglements, are used to study quantum information processes, in
particular, quantum computation, quantum teleportation, quantum cryptography 
\cite{Ben93,Eke,JS}. There have been mathematical studies of the
entanglements in \cite{Sch1,BBPSSW,Maj}, in which the entangled state is
defined by a compound state which can not be written as a convex combination 
$\sum_n\mu \left( n\right) \varsigma _n\otimes \varrho _n$ with any states $%
\varrho _n$ and $\varsigma _n.$ However it is obvious that there exist
several important applications with correlated states written as separable
forms above. Such correlated, or entangled states have been also discussed
in several contexts in quantum probability such as quantum measurement and
filtering \cite{Bel80, Bel94}, quantum compound state\cite{Ohy83,Ohy83-2}
and lifting \cite{ AcO}. In this paper, we study the mathematical structure
of quantum entangled states to provide a finer classification of quantum
sates, and we discuss the informational degree of entanglement and entangled
quantum mutual entropy.

We show that the pure entangled states can be treated as generalized
compound states, the nonseparable states of quantum compound systems which
are not representable by convex combinations of the product states.

The mixed compound states, defined as convex combinations by orthogonal
decompositions of their input marginal states $\varrho _0$, have been
introduced in \cite{Ohy83} for studying the information in a quantum channel
with the general output C*-algebra $\mathcal{A}$. This o-entangled compound
state is a particular case of so called separable state of a compound
system, the convex combination of the arbitrary product states which we call
c-entangled. We shall prove that the o-entangled compound states are most
informative among c-entangled states in the sense that the maximum of mutual
information over all c-entanglements to the quantum system $\left( \mathcal{A%
},\varrho \right) $ is achieved on the extreme o-entangled states, defined
by a Schatten decomposition of a given state $\varrho $ on $\mathcal{A}$.
This maximum coincides with von Neumann entropy $S\left( \varrho \right) $
of the state $\varrho $, and it can also be achieved as the maximum of the
mutual information over all couplings with classical probe systems described
by a maximal Abelian subalgebra $\mathcal{A}^{\circ }\subseteq \mathcal{A}$.
Thus the couplings described by c-entanglements of (quantum) probe systems $%
\mathcal{B}$ to a given system $\mathcal{A}$ don't give an advantage in
maximizing the mutual information in comparison with the quantum-classical
couplings, corresponding to the Abelian $\mathcal{B}=\mathcal{A}^{\circ }$.
The achieved maximal information $S\left( \varrho \right) $ coincides with
the classical entropy on the Abelian subalgebra $\mathcal{A}^{\circ }$ of a
Schatten decomposition for $\varrho $, and is bounded by $\ln \mathrm{rank}%
\mathcal{A}=\ln \dim \mathcal{A}^{\circ }$, where $\mathrm{rank}\mathcal{A}$
is the rank of the von Neumann algebra $\mathcal{A}$ defined as the
dimensionality of a maximal Abelian subalgebra. Due to $\dim \mathcal{A}\leq
\left( \mathrm{rank}\mathcal{A}\right) ^2$, it is achieved on the normal
central $\rho =\left( \mathrm{rank}\mathcal{A}\right) ^{-1}I$ only in the
case of finite dimensional $\mathcal{A}$.

More general than o-entangled states, the d-entangled states, are defined as
c-entangled states by orthogonal decomposition of only one marginal state on
the probe algebra $\mathcal{B}$. They can give bigger mutual entropy for a
quantum noisy channel than the o-entangled state which gains the same
information as d-entangled extreme states in the case of a deterministic
channel.

We prove that the truly (strongest) entangled states are most informative in
the sense that the maximum of mutual entropy over all entanglements to the
quantum system $\mathcal{A}$ is achieved on the quasi-compound state, given
by an extreme entanglement of the probe system $\mathcal{B}=\mathcal{A}$
with coinciding marginals, called standard for a given $\varrho $. The
standard entangled state is o-entangled only in the case of Abelian $%
\mathcal{A}$ or pure marginal state $\varrho $. The gained information for
such extreme q-compound state defines another type of entropy, the
quasi-entropy $S_q\left( \varrho \right) $ which is bigger than the von
Neumann entropy $S\left( \varrho \right) $ in the case of non-Abelian $%
\mathcal{A}$ (and mixed $\varrho $.) The maximum of mutual entropy over all
quantum couplings, described by true quantum entanglements of probe systems $%
\mathcal{B}$ to the system $\mathcal{A}$ is bounded by $\ln \mathrm{\dim }%
\mathcal{A}$, the logarithm of the dimensionality of the von Neumann algebra 
$\mathcal{A}$, which is achieved on a normal tracial $\rho $ in the case of
finite dimensional $\mathcal{A}$. Thus the q-entropy $S_q\left( \varrho
\right) $, which can be called the dimensional entropy, is the true quantum
entropy, in contrast to the von Neumann rank entropy $S\left( \varrho
\right) $, which is semi-classical entropy as it can be achieved as a
supremum over all couplings with the classical probe systems $\mathcal{B}$.
These entropies coincide in the classical case of Abelian $\mathcal{A}$ when 
$\mathrm{rank}\mathcal{A}=\dim \mathcal{A}$. In the case of non-Abelian
finite-dimensional $\mathcal{A}$ the q-capacity $C_q=\ln \mathrm{\dim }%
\mathcal{A}$ is achieved as the supremum of mutual entropy over all
q-encodings (correspondences), described by entanglements. It is strictly
bigger then the semi-classical capacity $C=\mathrm{\ln rank}\mathcal{A}$ of
the identity channel, which is achieved as the supremum over usual
encodings, described by the classical-quantum correspondences $\mathcal{A}%
^{\circ }\rightarrow \mathcal{A}$.

In this paper we consider the case of a discrete decomposable C*-algebra $%
\mathcal{A}$ for which the results are achieved by relatively simple proofs.
The purely quantum case of a simple algebra $\mathcal{A}=\mathcal{L}\left( 
\mathcal{H}\right) $, for which some proofs are rather obvious was
considered in a short paper \cite{BeO98}. The general case of decomposable
C*-algebra $\mathcal{A}$ to include the continuous systems, and will be
published elsewhere.

\section{Compound States and Entanglements}

Let $\mathcal{H}$ denote the (separable) Hilbert space of a quantum system,
and $\mathcal{L}\left( \mathcal{H}\right) $ be the algebra of all linear
bounded operators {on }$\mathcal{H}$. In order to include the classical
discrete systems as a particular quantum case, we shall fix a decomposable
subalgebra $\mathcal{A}\subseteq \mathcal{L}\left( \mathcal{H}\right) $ of
bounded observables $A\in \mathcal{A}$ of the form $A=\left[ A\left(
i\right) \delta _i^k\right] $, where $A\left( i\right) \in \mathcal{L}\left( 
\mathcal{H}_i\right) $ are arbitrary bounded operators in Hilbert subspaces $%
\mathcal{H}_i$, corresponding to an orthogonal decomposition $\mathcal{H}%
=\oplus _i\mathcal{H}_i$. A bounded linear functional $\varrho :\mathcal{{\ A%
}\rightarrow }\mathbf{C}$ is called a state on $\mathcal{A}$ if it is
positive (i.e., $\varrho \left( A\right) \geq 0$ for any positive operator $%
A $ in $\mathcal{A}$) and normalized $\varrho (I)=1$ for the identity
operator $I$ in $\mathcal{A}$ . A normal state can be expressed as 
\begin{equation}
\varrho \left( A\right) =\mathrm{tr}_{\mathcal{G}}\kappa ^{\dagger }A\kappa =%
\mathrm{tr}A\rho ,\text{ \quad }A\in \mathcal{A}  \label{1.1}
\end{equation}
where $\mathcal{G}$ is another separable Hilbert space, $\kappa $ is a
linear Hilbert-Schmidt operator from $\mathcal{G}$ to $\mathcal{H}$ and $%
\kappa ^{\dagger }$ is the adjoint operator of $\kappa $ from $\mathcal{H}$
to $\mathcal{G}$. This $\kappa $ is called the amplitude operator, and it is
called just the amplitude if $\mathcal{G}$ is one dimensional space $\Bbb{C}$%
, corresponding to the pure state $\varrho \left( A\right) =\kappa ^{\dagger
}A\kappa $ for a $\kappa \in \mathcal{H}$ with $\kappa ^{\dagger }\kappa
=\Vert \kappa \Vert ^2=1,$ in which case $\kappa ^{\dagger }$ is the adjoint
functional from $\mathcal{H}$ to $\Bbb{C}$. The density operator $\rho
=\kappa \kappa ^{\dagger }$ is uniquely defined by the condition $\rho \in 
\mathcal{A}$ as a decomposable trace class operator $\mathrm{P}_{\mathcal{A}%
}=\oplus \mathrm{P}_{\mathcal{A}}\left( i\right) $ with \textrm{P}$_{%
\mathcal{A}}\left( i\right) \in \mathcal{L}\left( \mathcal{H}_i\right) $, 
\begin{equation*}
\nu \left( i\right) =\mathrm{tr}_{\mathcal{H}_i}\mathrm{P}_{\mathcal{A}%
}\left( i\right) \geq 0,\quad \sum_i\nu \left( i\right) =1.
\end{equation*}
Thus the predual space $\mathcal{A}_{*}$ can be identified with the direct
sum $\oplus \mathcal{T}\left( \mathcal{H}_i\right) $ of the Banach spaces $%
\mathcal{T}\left( \mathcal{H}_i\right) $ of trace class operators in $%
\mathcal{H}_i$ (the density operators $\mathrm{P}_{\mathcal{A}}\in \mathcal{A%
}_{*}$, $\mathrm{P}_{\mathcal{B}}\in \mathcal{B}_{*}$ of the states $\varrho 
$, $\varsigma $ on different algebras $\mathcal{A}$, $\mathcal{B}$ will be
usually denoted by different letters $\rho ,\sigma $ corresponding to their
Greek variations $\varrho $, $\varsigma $.)

In general, $\mathcal{G}$ is not one dimensional, the dimensionality $\dim 
\mathcal{G}$ must be not less than $\mathrm{rank}\rho $, the dimensionality
of the range $\mathrm{ran}\rho \subseteq \mathcal{H}$ of the density
operator $\rho .$ We shall equip it with an isometric involution $%
J=J^{\dagger }$, $J^2=I$, having the properties of complex conjugation on $%
\mathcal{G}$, 
\begin{equation*}
J\sum \lambda _j\zeta _j=\sum \bar{\lambda _j}J\zeta _j,\quad \forall
\lambda _j\in \mathbf{C},\zeta _j\in \mathcal{G}
\end{equation*}
with respect to which $J\sigma =\sigma J$ for the positive and so
self-adjoint operator $\sigma =\kappa ^{\dagger }\kappa =\sigma ^{\dagger }$
on $\mathcal{G}$. The latter can also be expressed as the symmetricity
property $\tilde{\varsigma}=\varsigma $ of the state $\varsigma \left(
B\right) =$ $\mathrm{tr}B\sigma $ given by the real and so symmetric density
operator $\bar{\sigma}=\sigma =\tilde{\sigma}$ on $\mathcal{G}$ with respect
to the complex conjugation $\bar{B}=JBJ$ and the tilda operation ($\mathcal{G%
}$-transponation) $\tilde{B}=JB^{\dagger }J$ on the algebra $\mathcal{L}%
\left( \mathcal{G}\right) $, and thus on any tilda invariant decomposable
subalgebra $\mathcal{B}\subseteq \mathcal{L}\left( \mathcal{G}\right) $
containing $\kappa ^{\dagger }\mathcal{A}\kappa \ni \sigma $.

For example, $\mathcal{G}$ can be realized as a subspace of $l^2(\mathbf{N})$
of complex sequences $\mathbf{N}\ni n\mapsto \zeta \left( n\right) \in \Bbb{C%
}$, with $\sum_n\left| \zeta \left( n\right) \right| ^2<+\infty $ in the
diagonal representation $\sigma =\left[ \mu \left( n\right) \delta
_n^m\right] $. The involution $J$ can be identified with the complex
conjugation $C\zeta \left( n\right) =\bar{\zeta}\left( n\right) $, i.e., 
\begin{equation*}
C:\zeta =\sum_n|n\rangle \zeta \left( n\right) \mapsto C\zeta
=\sum_n|n\rangle \bar{\zeta}\left( n\right)
\end{equation*}
in the standard basis $\left\{ |n\rangle \right\} \subset \mathcal{G}$ of $%
l^2(\mathbf{N})$. In this case $\kappa =\sum \kappa _n\langle n|$ is given
by orthogonal eigen-amplitudes $\kappa _n\in \mathcal{H}$, $\kappa
_m^{\dagger }\kappa _n=0$, $m\neq n$, normalized to the eigen-values $%
\lambda \left( n\right) =\kappa _n^{\dagger }\kappa _n=\mu \left( n\right) $
of the density operator $\rho $ such that $\rho =\sum \kappa _n\kappa
_n^{\dagger }$ is a Schatten decomposition, i.e. the spectral decomposition
of $\rho $ into one-dimensional orthogonal projectors. In any other basis
the operator $J$ is defined then by $J=U^{\dagger }CU$, where $U$ is the
corresponding unitary transformation. One can also identify $\mathcal{G}$
with $\mathcal{H}$ by $U\kappa _n=\lambda \left( n\right) ^{1/2}|n\rangle $
such that the operator $\rho $ is real and symmetric, $J\rho J=\rho =J\rho
^{\dagger }J$ in $\mathcal{G}=\mathcal{H}$ with respect to the involution $J$
defined in $\mathcal{H}$ by $J\kappa _n=\kappa _n$. Here $U$ is an isometric
operator $\mathcal{H}\rightarrow l^2\left( \Bbb{N}\right) $ diagonalizing
the operator $\rho $: $U\rho U^{\dagger }=\sum |n\rangle \lambda \left(
n\right) \langle n|$. The amplitude operator $\kappa =\rho ^{1/2}$
corresponding to $\mathcal{B}=\mathcal{A}$, $\sigma =\rho $ is called
standard.

Given the amplitude operator $\kappa $, one can define not only the states $%
\varrho $ and $\varsigma $ on the algebras $\mathcal{A}=\mathcal{L}\left( 
\mathcal{H}\right) $ and $\mathcal{B}=\mathcal{L}\left( \mathcal{G}\right) $
but also a pure entanglement state $\varpi $ on the algebra $\mathcal{B}%
\otimes \mathcal{A}$ of all bounded operators on the tensor product Hilbert
space $\mathcal{G}\otimes \mathcal{H}$ by

\begin{equation*}
\varpi \left( B\otimes A\right) =\mathrm{tr}_{\mathcal{G}}\tilde{B}\kappa
^{\dagger }A\kappa =\mathrm{tr}_{\mathcal{H}}A\kappa \tilde{B}\kappa
^{\dagger }.
\end{equation*}
Indeed, thus defined $\varpi $ is uniquely extended by linearity to a normal
state on the algebra $\mathcal{B}\otimes \mathcal{A}$ generated by all
linear combinations $C=\sum \lambda _jB_j\otimes A_j$ due to $\varpi \left(
I\otimes I\right) =\mathrm{tr}\kappa ^{\dagger }\kappa =1$ and 
\begin{eqnarray*}
\varpi \left( C^{\dagger }C\right) &=&\sum_{i,k}\bar{\lambda}_i\lambda _k%
\mathrm{tr}_{\mathcal{G}}\tilde{B}_k\tilde{B}_i^{\dagger }\kappa ^{\dagger
}A_i^{\dagger }A_k\kappa \\
&=&\sum_{i,k}\bar{\lambda}_i\lambda _k\mathrm{tr}_{\mathcal{G}}\tilde{B}%
_i^{\dagger }\kappa ^{\dagger }A_i^{\dagger }A_k\kappa \tilde{B}_k=\mathrm{tr%
}_{\mathcal{G}}\chi ^{\dagger }\chi \geq 0,
\end{eqnarray*}
where $\chi =\sum_jA_j\kappa \tilde{B}_j$. This state is pure on $\mathcal{L}%
\left( \mathcal{G}\otimes \mathcal{H}\right) $ as it is given by an
amplitude $\vartheta \in \mathcal{G}\otimes \mathcal{H}$ defined as 
\begin{equation*}
\left( \zeta \otimes \eta \right) ^{\dagger }\vartheta =\eta ^{\dagger
}\kappa J\zeta ,\quad \forall \zeta \in \mathcal{G},\eta \in \mathcal{H},
\end{equation*}
and it has the states $\varrho $ and $\varsigma $ as the marginals of $%
\varpi $: 
\begin{equation}
\varpi \left( I\otimes A\right) =\mathrm{tr}_{\mathcal{H}}A\rho ,\quad
\varpi \left( B\otimes I\right) =\mathrm{tr}_{\mathcal{G}}B\sigma .
\label{1.2}
\end{equation}
As follows from the next theorem for the case $\mathcal{F}=\Bbb{C}$ , any
pure state 
\begin{equation*}
\varpi \left( B\otimes A\right) =\vartheta ^{\dagger }\left( B\otimes
A\right) \vartheta ,\quad B\in \mathcal{B},A\in \mathcal{A}
\end{equation*}
given on $\mathcal{L}\left( \mathcal{G}\otimes \mathcal{H}\right) $ by an
amplitude $\vartheta \in \mathcal{G}\otimes \mathcal{H}$ with $\vartheta
^{\dagger }\vartheta =1$, can be achieved by a unique entanglement of its
marginal states $\varsigma $ and $\varrho $.\smallskip

\bigskip

\noindent
{\bf Theorem 2.1.} 
{\it
Let $\varpi :\mathcal{B}\otimes \mathcal{A}\rightarrow \Bbb{C}$ be a
compound state 
\begin{equation*}
\varpi \left( B\otimes A\right) =\mathrm{tr}_{\mathcal{F}}\upsilon ^{\dagger
}\left( B\otimes A\right) \upsilon ,
\end{equation*}
defined by an amplitude operator $\upsilon :\mathcal{F}\rightarrow \mathcal{G%
}\otimes \mathcal{H}$ on a separable Hilbert space $\mathcal{F}$ into the
tensor product Hilbert space $\mathcal{G}\otimes \mathcal{H}$ with 
\begin{equation*}
\upsilon \upsilon ^{\dagger }\in \mathcal{B}\otimes \mathcal{A},\quad 
\mathrm{tr}_{\mathcal{F}}\upsilon ^{\dagger }\upsilon =1.
\end{equation*}
Then this state can be achieved as an entanglement 
\begin{equation}
\varpi \left( B\otimes A\right) =\mathrm{tr}_{\mathcal{G}}\tilde{B}\kappa
^{\dagger }\left( I\otimes A\right) \kappa =\mathrm{tr}_{\mathcal{F}\otimes 
\mathcal{H}}\left( I\otimes A\right) \kappa \tilde{B}\kappa ^{\dagger }
\label{1.4}
\end{equation}
of the states (\ref{1.2}) with $\sigma =\kappa ^{\dagger }\kappa $ and $\rho
=\mathrm{tr}_{\mathcal{F}}\kappa \kappa ^{\dagger }$, where $\kappa $ is an
amplitude operator $\mathcal{G}\rightarrow \mathcal{F}\otimes \mathcal{H}$
with 
\begin{equation}
\kappa ^{\dagger }\left( I\otimes \mathcal{A}\right) \kappa \subset \mathcal{%
B},\;\mathrm{tr}_{\mathcal{F}}\kappa \mathcal{B}\kappa ^{\dagger }\subset 
\mathcal{A}.  \label{1.3}
\end{equation}
The entangling operator $\kappa $ is uniquely defined by $\tilde{\kappa}%
U=\upsilon $ up to a unitary transformation $U$ of the minimal domain $%
\mathcal{F}=\mathrm{dom}\upsilon $.
}

\bigskip

\proof %
Without loss of generality we can assume that the space $\mathcal{F}$ is
equipped with an isometric involution $J$ as well as the space $\mathcal{G}$
is equipped with $J$ . The entangling operator $\kappa $ can be defined then
as $\kappa =\left( U\otimes I\right) \tilde{\upsilon}$ by 
\begin{equation*}
(U\xi \otimes \eta )^{\dagger }\kappa \zeta =(\xi \otimes \eta )^{\dagger }%
\tilde{\upsilon}\zeta :=\left( J\zeta \otimes \eta \right) ^{\dagger
}\upsilon J\xi ,\quad \forall \xi \in \mathcal{F},\zeta \in \mathcal{G},\eta
\in \mathcal{H},
\end{equation*}
where $U$ is arbitrary linear isometry in $\mathcal{F}$. Indeed, let $%
\left\{ \xi _k\right\} $ be an orthonormal basis of $\mathcal{F}$\textbf{, }%
in which,\textbf{\ }say (but not necessary,) the density operator $\upsilon
^{\dagger }\upsilon $ is diagonal, and $J:\mathcal{F}\rightarrow \mathcal{F}$
be the complex conjugation in this basis, $J\xi _k=\xi _k$, defining an
isometric involution in $\mathcal{F}$. In general $J$ is different from the
complex conjugation $C$, given by $C|n\rangle =|n\rangle $ in the standard
basis $\left\{ |n\rangle ;n\in \Bbb{N}\right\} $ if $\mathcal{F}$ is
identified with a subspace $l^2\left( \mathbf{N}\right) $ for the diagonal
representation of $\upsilon ^{\dagger }\upsilon $. Note that although the
isometric transformation $U=\sum_k|k\rangle \xi _k^{\dagger }$ of the
arbitrary basis $\left\{ \xi _k\right\} \subset \mathcal{F}$ into $\left\{
|k\rangle \right\} \subset l^2\left( \mathbf{N}\right) $ is also arbitrary,
it can be always considered as real with respect to $C$ and $J=U^{\dagger
}CU $, in the sense $\bar{U}:=CUJ=U$, and so $\tilde{U}:=CU^{\dagger
}J=U^{\dagger }$. Defining $\kappa =\sum \kappa _n\langle n|$ in the
standard basises of $\mathcal{F}$ and $\mathcal{G}$ as the block-matrix $%
\sum_{kn}|k\rangle \otimes \psi _k\left( n\right) \langle n|$ transposed to $%
\sum_{kn}|n\rangle \otimes \psi _k\left( n\right) \langle k|$, where the
amplitudes $\psi _k\left( n\right) \in \mathcal{H}$ are given by the matrix
elements $\eta ^{\dagger }\psi _k\left( n\right) =\left( \langle n|\otimes
\eta ^{\dagger }\right) \upsilon \xi _k$, we obtain 
\begin{eqnarray*}
\mathrm{tr}_{\mathcal{G}}\tilde{B}\kappa ^{\dagger }\left( I\otimes A\right)
\kappa &=&\sum_{n,m}\left\langle n\right| \tilde{B}\left| m\right\rangle
\psi _k^{\dagger }\left( m\right) A\psi _k\left( n\right) \\
&=&\sum_{n,m}\psi _k^{\dagger }\left( m\right) \left\langle m\right| B\left|
n\right\rangle A\psi _k\left( n\right) =\mathrm{tr}_{\mathcal{F}}\upsilon
^{\dagger }\left( B\otimes A\right) \upsilon \text{ .}
\end{eqnarray*}
Hence $\kappa :\mathcal{G}\rightarrow \mathcal{F}\otimes \mathcal{H}$ ,
defined by $\kappa _n=\sum |k\rangle \psi _k\left( n\right) $ as the
transposed to $\upsilon U^{\dagger }=\upsilon \tilde{U}\equiv \tilde{\kappa}$%
, is the required entangling operator of the form $\kappa =\left( U\otimes
I\right) \tilde{\upsilon}$ with $\Bbb{\kappa }^{\dagger }\Bbb{\kappa }$$%
=\sigma =\mathrm{tr}_{\mathcal{H}}\upsilon \upsilon ^{\dagger }$ and $%
\mathrm{tr}_{\mathcal{F}}\kappa \kappa ^{\dagger }=\rho =\mathrm{tr}_{%
\mathcal{G}}\upsilon \upsilon ^{\dagger }$. Moreover, it satisfies the
conditions (\ref{1.3}) as $\omega =\upsilon \upsilon ^{\dagger }\in \mathcal{%
B}\otimes \mathcal{A}$ and thus 
\begin{equation*}
\kappa ^{\dagger }\left( I\otimes A\right) \kappa =\mathrm{tr}_{\mathcal{H}%
}\left( I\otimes A\right) \omega \in \mathcal{B},\;\mathrm{tr}_{\mathcal{F}%
}\kappa \tilde{B}\kappa ^{\dagger }=\mathrm{tr}_{\mathcal{G}}\left( B\otimes
I\right) \omega \in \mathcal{A}.
\end{equation*}

The uniqueness follows from the obvious isometricity of the families 
\begin{equation*}
\left\{ \sum_k|k\rangle \eta ^{\dagger }\psi _k\left( n\right) :n\in \Bbb{N}%
,\eta \in \mathcal{H}\right\} ,\quad \left\{ \sum_k\eta ^{\dagger }\psi
_k\left( n\right) \xi _k^{\dagger }:n\in \Bbb{N},\eta \in \mathcal{H}\right\}
\end{equation*}
of vectors $\left( I\otimes \eta ^{\dagger }\right) \kappa |n\rangle $ in $%
\mathcal{F}\subseteq l^2\left( \mathbf{N}\right) $ and of $\left( \langle
n|\otimes \eta ^{\dagger }\right) \upsilon $ in $\mathcal{F}^{\dagger }$
which follows from 
\begin{equation*}
\mathrm{tr}_{\mathcal{G}}|n\rangle \langle m|\kappa ^{\dagger }\left(
I\otimes \eta \eta ^{\dagger }\right) \kappa =\mathrm{tr}_{\mathcal{F}%
}\upsilon ^{\dagger }\left( |m\rangle \langle n|\otimes \eta \eta ^{\dagger
}\right) \upsilon .
\end{equation*}
Thus they are unitary equivalent in the minimal space $\mathcal{F}$. So the
entangling operator $\kappa $ is defined in the minimal $\mathcal{F}$ up to
the unitary equivalence, corresponding to the arbitrary of the unitary
operator $U$ in $\mathcal{F}$, intertwining the involutions $C$ and $J$.%
\endproof %

Note that the entangled state (\ref{1.4}) is written as 
\begin{equation*}
\varpi \left( B\otimes A\right) =\mathrm{tr}_{\mathcal{G}}\tilde{B}\pi
\left( A\right) =\mathrm{tr}_{\mathcal{H}}A\pi _{*}\left( \tilde{B}\right) ,
\end{equation*}
where $\pi \left( A\right) =\kappa ^{\dagger }\left( I\otimes A\right)
\kappa $, bounded by $\left\| A\right\| \sigma \in \mathcal{B}_{*}$ for any $A\in \mathcal{L}\left( \mathcal{H}\right) $, is in the predual space $\mathcal{B}_{*}\subset \mathcal{B}$ of all trace-class operators in $%
\mathcal{G}$, and $\pi _{*}\left( B\right) =\mathrm{tr}_{\mathcal{F}}\kappa
B\kappa ^{\dagger }$, bounded by $\left\| B\right\| \rho \in \mathcal{A}_{*}$, is in $\mathcal{A}_{*}\subset \mathcal{A}$. The map $\pi $ is the
Steinspring form \cite{Sti55} of the general completely positive map $%
\mathcal{A}\rightarrow \mathcal{B}_{*}$, written in the eigen-basis $\left\{
\left| k\right\rangle \right\} \subset \mathcal{F}$ of the density operator $\upsilon ^{\dagger }\upsilon $ as 
\begin{equation}
\pi \left( A\right) =\sum_{m,n}\left| m\right\rangle \kappa _m^{\dagger
}\left( I\otimes A\right) \kappa _n\left\langle n\right| ,\quad A\in 
\mathcal{A}  \label{1.7}
\end{equation}
while the dual operation $\pi _{*}$ is the Kraus form \cite{Kra71} of the
general completely positive map $\mathrm{A}\rightarrow \mathcal{A}_{*}$,
given in this basis as 
\begin{equation}
\pi _{*}\left( B\right) =\sum_{n,m}\left\langle n\right| B\left|
m\right\rangle \mathrm{tr}_{\mathcal{F}}\kappa _n\kappa _m^{\dagger }=%
\mathrm{tr}_{\mathcal{G}}\tilde{B}\omega .  \label{1.8}
\end{equation}
It corresponds to the general form 
\begin{equation}
\omega =\sum_{m,n}|n\rangle \langle m|\otimes \mathrm{tr}_{\mathcal{F}%
}\kappa _n\kappa _m^{\dagger }  \label{1.9}
\end{equation}
of the density operator $\omega =\upsilon \upsilon ^{\dagger }$ for the
entangled state $\varpi \left( B\otimes A\right) =\mathrm{tr}\left( B\otimes
A\right) \omega $ in this basis, characterized by the weak orthogonality
property 
\begin{equation}
\mathrm{tr}_{\mathcal{F}}\psi \left( m\right) ^{\dagger }\psi \left(
n\right) =\mu \left( n\right) \delta _n^m  \label{1.10}
\end{equation}
in terms of the amplitude operators $\psi \left( n\right) =\left( I\otimes
\langle n|\right) \tilde{\kappa}=\tilde{\kappa}_n$.

\bigskip

\noindent
{\bf Definition 2.1.} 
{\it
The dual map $\pi _{*}:\mathcal{B}\rightarrow \mathcal{A}_{*}$ to a
completely positive map $\pi :\mathcal{A}\rightarrow \mathcal{B}_{*}$,
normalized as $\mathrm{tr}_{\mathcal{G}}\pi \left( I\right) =1$, is called
the quantum entanglement of the state $\varsigma =\pi \left( I\right) $ on $%
\mathcal{B}$ to the state $\varrho =\pi _{*}\left( I\right) $ on $\mathcal{A}
$. The entanglement by 
\begin{equation}
\pi _{*}^{\circ }\left( A\right) =\rho ^{1/2}A\rho ^{1/2}=\pi ^{\circ
}\left( A\right)  \label{1.5}
\end{equation}
of the state $\varsigma =\varrho $ on the algebra $\mathcal{B}=\mathcal{A}$
is called standard for the system $\left( \mathcal{A},\varrho \right) $.
}

\bigskip

The standard entanglement defines the standard compound state 
\begin{equation*}
\varpi _0\left( B\otimes A\right) =\mathrm{tr}_{\mathcal{H}}\tilde{B}\rho
^{1/2}A\rho ^{1/2}=\mathrm{tr}_{\mathcal{H}}A\rho ^{1/2}\tilde{B}\rho ^{1/2}
\end{equation*}
on the algebra $\mathcal{A}\otimes \mathcal{A}$, which is pure, given by the
amplitude $\vartheta _0=\tilde{\kappa}_0$, where $\kappa _0=\rho ^{1/2}$ in
the case of the simple algebra $\mathcal{A}=\mathcal{L}\left( \mathcal{H}%
\right) $. In the general case of decomposable $\mathcal{A}=\oplus \mathcal{L%
}\left( \mathcal{H}_i\right) $ with the density operator $\rho =\oplus \rho
\left( i\right) $ having more than one components $\rho \left( i\right)
=\rho _i\nu \left( i\right) $ with $\nu \left( i\right) =\mathrm{tr}\rho
\left( i\right) \neq 0$ and positive $\rho _i\in \mathcal{L}\left( \mathcal{H%
}_i\right) $, the standard state $\varpi _0$ is a mixture 
\begin{equation}
\varpi _0\left( B\otimes A\right) =\sum_i\vartheta _0^{i\dagger }\left(
B\left( i\right) \otimes A\left( i\right) \right) \vartheta _0^i\nu \left(
i\right) ,\quad A,B\in \mathcal{A}  \label{1.6}
\end{equation}
of such pure compound states given by the amplitudes $\vartheta _0^i\in 
\mathcal{H}_i\otimes \mathcal{H}_i$ with $\tilde{\vartheta}_0^i\tilde{%
\vartheta}_0^{i\dagger }=\rho _i$. The standard amplitudes $\vartheta
_0^i\in \mathcal{H}_i\otimes \mathcal{H}_i$ for an orthogonal decomposition $%
\upsilon _0=\sum_i\vartheta _0^i\xi _i^{\dagger }\nu \left( i\right) ^{1/2}$
of the standard amplitude operator $\upsilon _0:\mathcal{F}_0\rightarrow 
\mathcal{H}\otimes \mathcal{H}$ are defined as $\tilde{\kappa}_0\left(
i\right) /\left\| \tilde{\kappa}_0\left( i\right) \right\| $ by the
entangling components $\kappa _0\left( i\right) =\rho \left( i\right) ^{1/2}$
with 
\begin{equation*}
\left( \zeta _i\otimes \eta _i\right) ^{\dagger }\tilde{\kappa}_0\left(
i\right) =\eta _i^{\dagger }\kappa _0\left( i\right) J\zeta _i,\quad \forall
\eta _i,\zeta _i\in \mathcal{H}_i.
\end{equation*}

[Example] In quantum physics the entangled states are usually obtained by a
unitary transformation $U$ of an initial disentangled state, described by
the density operator $\sigma _0\otimes \rho _0\otimes \tau _0$ on the tensor
product Hilbert space $\mathcal{G}\otimes \mathcal{H}\otimes \mathcal{K}$ ,
that is, 
\begin{equation*}
\varpi \left( B\otimes A\right) =\mathrm{tr}U_1^{\dagger }\left( B\otimes
A\otimes I\right) U_1\left( \sigma _0\otimes \rho _0\otimes \tau _0\right) .
\end{equation*}
In the simple case, when $\mathcal{K}=\Bbb{C}$, $\tau _0=1$, the joint
amplitude operator $\upsilon $ is defined on the tensor product $\mathcal{F}=%
\mathcal{G}\otimes \mathcal{H}_0$ with $\mathcal{H}_0=\mathrm{ran}\rho _0$
as $\upsilon =U_1\left( \sigma _0\otimes \rho _0\right) ^{1/2}$. The
entangling operator $\kappa $, describing the entangled state $\varpi $, is
constructed as it was done in the proof of Theorem 1 by transponation of the
operator $\upsilon U^{\dagger }$, where $U$ is arbitrary isometric operator $%
\mathcal{F}\rightarrow \mathcal{G}\otimes \mathcal{H}_0$. The dynamical
procedure of such entanglement in terms of the completely positive map $\pi
_{*}:\mathcal{A}\rightarrow \mathcal{B}_{*}$ is the subject of Belavkin
quantum filtering theory \cite{Bel97}. The quantum filtering dilation
theorem \cite{Bel97} proves that any entanglement $\pi $ can be obtained the
unitary entanglement as the result of quantum filtering by tracing out some
degrees of freedom of a quantum environment, described by the density
operator $\tau _0$ on the Hilbert space $\mathcal{K}$, even in the
continuous time case.

\section{C- and D-Entanglements and Encodings}

The compound states play the role of joint input-output probability measures
in classical information channels, and can be pure in quantum case even if
the marginal states are mixed. The pure compound states achieved by an
entanglement of mixed input and output states exhibit new, non-classical
type of correlations which are responsible for the EPR type paradoxes in the
interpretation of quantum theory. The mixed compound states on $\mathcal{B}%
\otimes \mathcal{A}$ which are given as the convex combinations 
\begin{equation*}
\varpi =\sum_n\varsigma _n\otimes \varrho _n\mu \left( n\right) ,\quad \mu
\left( n\right) \geq 0,\;\sum_n\mu \left( n\right) =1
\end{equation*}
of tensor products of pure or mixed normalized states $\varrho _n\in 
\mathcal{A}_{*}$, $\varsigma _n\in \mathcal{B}_{*}$ as in classical case, do
not exhibit such paradoxical behavior, and are usually considered as the
proper candidates for the input-output states in the communication channels.
Such separable compound states are achieved by c-entanglements, the convex
combinations of the primitive entanglements $B\mapsto \mathrm{tr}_{\mathcal{G%
}}B\omega _n$, given by the density operators $\omega _n=\sigma _n\otimes
\rho _n$ of the product states $\varpi _n=\varsigma _n\otimes \varrho _n$: 
\begin{equation}
\pi _{*}\left( B\right) =\sum_n\rho _n\mathrm{tr}_{\mathcal{G}}B\sigma _n\mu
\left( n\right) ,  \label{2.1}
\end{equation}
A compound state of this sort was introduced by Ohya \cite{Ohy83} in order
to define the quantum mutual entropy expressing the amount of information
transmitted from an input quantum system to an output quantum system through
a quantum channel, using a Schatten decomposition $\sigma =\sum_n\sigma
_n\mu \left( n\right) $, $\sigma _n=|n\rangle \langle n|$ of the input
density operator $\sigma $. It corresponds to a particular, diagonal type 
\begin{equation}
\pi \left( A\right) =\sum_n|n\rangle \kappa _n^{\dagger }\left( I\otimes
A\right) \kappa _n\langle n|  \label{2.4}
\end{equation}
of the entangling map (\ref{1.7}) in an eigen-basis $\left\{ |n\rangle
\right\} \in \mathcal{G}$ of the density operator $\sigma $, and is
discussed in this section.

Let us consider a finite or infinite input system indexed by the natural
numbers $n\in \mathbf{N}$. The associated space $\mathcal{G}\subseteq
l^2\left( \mathbf{N}\right) $ is the Hilbert space of the input system
described by a quantum projection-valued measure $n\mapsto |n\rangle \langle
n|$ on $\Bbb{N}$, given an orthogonal partition of unity $I=\sum |n\rangle
\langle n|$ $\in \mathcal{B}$ of the finite or infinite dimensional input
Hilbert space $\mathcal{G}$. Each input pure state, identified with the
one-dimensional density operator $|n\rangle \langle n|\in \mathcal{B}$
corresponding to the elementary symbol $n\in \Bbb{N}$, defines the
elementary output state $\varrho _n$ on $\mathcal{A}$. If the elementary
states $\varrho _n$ are pure, they are described by output amplitudes $\eta
_n\in \mathcal{H}$ satisfying $\eta _n^{\dagger }\eta _n=1=\mathrm{tr}\rho
_n $, where $\rho _n=$ $\eta _n\eta _n^{\dagger }$ are the corresponding
output one-dimensional density operators. If these amplitudes are
non-orthogonal $\eta _m^{\dagger }\eta _n\neq \delta _n^m$, they cannot be
identified with the input amplitudes $|n\rangle $.

The elementary joint input-output states are given by the density operators $%
|n\rangle \langle n|\otimes \rho _n$ in $\mathcal{G}\otimes \mathcal{H}$.
Their mixtures 
\begin{equation}
\omega =\sum_n\mu \left( n\right) |n\rangle \langle n|\otimes \rho _n,
\label{2.2}
\end{equation}
define the compound states on $\mathcal{B}\otimes \mathcal{A}$, given by the
quantum correspondences $n\mapsto |n\rangle \langle n|$ with the
probabilities $\mu \left( n\right) $. Here we note that the quantum
correspondence is described by a classical-quantum channel, and the general
d-compound state for a quantum-quantum channel in quantum communication can
be obtained in this way due to the orthogonality of the decomposition (\ref
{2.2}), corresponding to the orthogonality of the Schatten decomposition $%
\sigma =\sum_n|n\rangle \mu \left( n\right) \langle n|$ for $\sigma =\mathrm{%
tr}_{\mathcal{H}}\omega $.

The comparison of the general compound state (\ref{1.9}) with (\ref{2.2})
suggests that the quantum correspondences are described as the diagonal
entanglements 
\begin{equation}
\pi _{*}\left( B\right) =\sum_n\mu \left( n\right) \langle n|B|n\rangle \rho
_n,  \label{2.3}
\end{equation}
They are dual to the orthogonal decompositions (\ref{2.4}): 
\begin{equation*}
\pi \left( A\right) =\sum_n\mu \left( n\right) |n\rangle \eta _n^{\dagger
}A\eta _n\langle n|=\sum_n|n\rangle \eta \left( n\right) ^{\dagger }A\eta
\left( n\right) \langle n|,
\end{equation*}
\newline
where $\eta \left( n\right) =\mu \left( n\right) ^{1/2}\eta _n$. These are
the entanglements with the stronger orthogonality 
\begin{equation}
\psi \left( m\right) \psi \left( n\right) ^{\dagger }=\rho \left( n\right)
\delta _n^m,  \label{2.5}
\end{equation}
for the amplitude operators $\psi \left( n\right) :\mathcal{F}\rightarrow 
\mathcal{H}$ of the decomposition $\upsilon =\sum_n|n\rangle \otimes \psi
\left( n\right) $ in comparison with the orthogonality (\ref{1.10}). The
orthogonality (\ref{2.5}) can be achieved in the following manner: Take in (%
\ref{1.7}) $\kappa _n=|n\rangle \otimes \eta \left( n\right) $ with $\langle
m|n\rangle =\delta _n^m$ so that 
\begin{equation*}
\kappa _m^{\dagger }\left( I\otimes A\right) \kappa _n=\mu \left( n\right)
\eta _n^{\dagger }A\eta _n\delta _n^m
\end{equation*}
for any $A\in \mathcal{A}$. Then the strong orthogonality condition (\ref
{2.5}) is fulfilled by the amplitude operators $\psi \left( n\right) =\eta
\left( n\right) \langle n|=\tilde{\kappa}_n$, and 
\begin{equation*}
\kappa ^{\dagger }\kappa =\sum_n\mu \left( n\right) |n\rangle \langle
n|=\sigma ,\;\kappa \kappa ^{\dagger }=\sum_n\eta \left( n\right) \eta
\left( n\right) ^{\dagger }=\rho .
\end{equation*}
It corresponds to the amplitude operator for the compound state (\ref{2.2})
of the form 
\begin{equation}
\upsilon =\sum_n\left| n\right\rangle \otimes \psi \left( n\right) U,
\label{2.6}
\end{equation}
where $U$ is arbitrary unitary operator from $\mathcal{F}$ onto $\mathcal{G}$%
, i.e. $\upsilon $ is unitary equivalent to the diagonal amplitude operator 
\begin{equation*}
\kappa =\sum_n|n\rangle \langle n|\otimes \eta \left( n\right)
\end{equation*}
on $\mathcal{F}=\mathcal{G}$ into $\mathcal{G}\otimes \mathcal{H}$. Thus, we
have proved the following theorem in the case of pure output states $\rho
_n=\eta _n\eta _n^{\dagger }$.

\bigskip

\noindent
{\bf Theorem 3.1.} 
{\it
Let $\pi $ be the operator (\ref{2.2}), defining a d-compound state of the
form 
\begin{equation}
\varpi \left( B\otimes A\right) =\sum_n\langle n|B|n\rangle \mathrm{tr}_{%
\mathcal{F}_n}\psi _n^{\dagger }A\psi _n\mu \left( n\right)  \label{2.7}
\end{equation}
Then it corresponds to the entanglement by the orthogonal decomposition (\ref
{2.4}) mapping the algebra $\mathcal{A}$ into a diagonal subalgebra of $%
\mathcal{B}$.
}

\bigskip

\proof %
Let $\oplus _n\mathcal{F}_n$ be the Hilbert orthogonal sum of the domains $%
\mathcal{F}_n$ for the amplitude operators $\psi _n$ in (\ref{2.7}) with an
isometric involution $\oplus C_n$. In the case $\mathcal{F}_n=\Bbb{C}$ of
the amplitudes $\psi _n\in \mathcal{H}$ corresponding to pure states $\rho
_n $ the involution $\oplus _nC_n$ is the componentwise complex conjugation
in $\oplus _n\Bbb{C}\subseteq l^2\left( \Bbb{N}\right) $; in the general
case it is given by some isometric involutions $C_n$ in the Hilbert spaces $%
\mathcal{F}_n$, which are equivalent to the ranges $\mathcal{H}_n=\rho _n%
\mathcal{H}$ of the density operators $\rho _n=\psi _n\psi _n^{\dagger }$
with the standard involutions in their eigen-representations, or contain
these ranges. We can define the global output amplitude operator $\psi
\left( n\right) $ on $\mathcal{F}=\oplus _n\mathcal{F}_n$ by 
\begin{equation*}
\psi \left( n\right) =\mu \left( n\right) ^{1/2}\psi _n\epsilon _n^{\dagger
},
\end{equation*}
where $\epsilon _n:\mathcal{F}_n\rightarrow \mathcal{F}$ are the canonical
orthogonal isometries, $\epsilon _m^{\dagger }\epsilon _n=I_n\delta _n^m$,
and by (\ref{2.6}) an amplitude operator $\upsilon :\mathcal{F}\rightarrow 
\mathcal{G}\otimes \mathcal{H}$ of the compound state (\ref{2.7}), defining
its density operator $\omega =\upsilon \upsilon ^{\dagger }$ independently
of the unitary transformation $U$ of the Hilbert space onto $\oplus _n%
\mathcal{F}_n $.

The entangling operator $\kappa =\sum_n\kappa _n\langle n|$ is then defined
by its components $\kappa _n\in \mathcal{F}\otimes \mathcal{H}$ of the form 
\begin{equation*}
\kappa _n=\left( \epsilon _n\otimes I\right) \tilde{\psi}_n\mu \left(
n\right) ^{1/2}=\tilde{\psi}\left( n\right) ,
\end{equation*}
Here $\tilde{\psi}_n$ are the amplitudes in $\mathcal{F}_n\otimes \mathcal{H}
$ obtained from the operators $\psi _n:\mathcal{F}_n\rightarrow \mathcal{H}$
by 
\begin{equation*}
\left( \xi _n\otimes \eta \right) ^{\dagger }\tilde{\psi}_n=\eta ^{\dagger
}\psi _nC_n\xi _n,\quad \forall \eta \in \mathcal{H},\xi _n\in \mathcal{F}_n
\end{equation*}
In particular $\kappa $ is the diagonal amplitude operator with the
components $\kappa _n=\oplus _m\delta _n^m\tilde{\psi}\left( n\right) $ in $%
\oplus _m\mathcal{F}_m\otimes \mathcal{H}$: 
\begin{equation}
\kappa =\sum_n\kappa _n\langle n|=\oplus _m\tilde{\psi}\left( m\right)
\langle m|.  \label{2.8}
\end{equation}
Thus the entanglement (\ref{1.8}) corresponding to (\ref{2.7}) is given by
the dual to (\ref{2.4}) diagonal map (\ref{2.3}) with the density operators $%
\rho \left( n\right) =\psi \left( n\right) \psi \left( n\right) ^{\dagger }=%
\mathrm{tr}_{\mathcal{F}}\kappa _n\kappa _n^{\dagger }$ normalized to the
probabilities $\mu \left( n\right) =\kappa _n^{\dagger }\kappa _n$.%
\endproof %

Note that (2.9) defines the general form of a positive map on $\mathcal{A}$
with values in the simultaneously diagonal trace-class operators in 
$\mathrm{A}$.

\bigskip

\noindent
{\bf Definition 3.1.} 
{\it
A convex combination (\ref{2.1}) of the primitive CP maps $\rho _n\varsigma
_n$ is called c-entanglement, and is called d-entanglement, or quantum
encoding if it has the diagonal form (\ref{2.3}) on $\mathcal{B}$. The
d-entanglement is called o-entanglement and compound state is called
o-compound if all density operators $\rho _n$ are orthogonal: $\rho _m\rho
_n=\rho _n\rho _m$ for all $m$ and $n$.
}

\bigskip

Note that due to the commutativity of the operators $B\otimes I$ with $%
I\otimes A$ on $\mathcal{G}\otimes \mathcal{H}$, one can treat the
correspondences as the nondemolition measurements \cite{Bel94} in $\mathcal{B%
}$ with respect to $\mathcal{A}$. So, the compound state is the state
prepared for such measurements on the input $\mathcal{G}$. It coincides with
the mixture of the states, corresponding to those after the measurement
without reading the sent message. The set of all d-entanglements
corresponding to a given Schatten decomposition of the input state $\sigma $
on $\mathcal{B}$ is obviously convex with the extreme points given by the
pure output states $\rho _n$ on $\mathcal{A}$, corresponding to a not
necessarily orthogonal decompositions $\rho =\sum_n\rho \left( n\right) $
into one-dimensional density operators $\rho \left( n\right) =\mu \left(
n\right) \rho _n.$

The Schatten decompositions $\rho =\sum_n\lambda \left( n\right) \rho _n$
correspond to the extreme d-entanglements, $\rho _n=\eta _n\eta _n^{\dagger
} $, $\mu \left( n\right) =\lambda \left( n\right) $, characterized by
orthogonality $\rho _m\rho _n=0$, $m\neq n$ . They form a convex set of
d-entanglements with mixed commuting $\rho _n$ for each Schatten
decomposition of $\rho $. The orthogonal d-entanglements were used in \cite
{AOW96} to construct a particular type of Accardi's transitional
expectations \cite{Acc74} and to define the entropy in a quantum dynamical
system via such transitional expectations.

The established structure of the general q-compound states suggests also the
general form 
\begin{equation*}
\Phi _{*}\left( B,\varrho _0\right) =\mathrm{tr}_{\mathcal{F}_1}X^{\dagger
}\left( B\otimes \rho _0\right) X=\mathrm{tr}_{\mathcal{G}}\left( \tilde{B}%
\otimes I\right) Y\left( I\otimes \rho _0\right) Y^{\dagger }
\end{equation*}
of transitional expectations $\Phi _{*}:\mathcal{B}\times \mathcal{A}%
_{*}^{\circ }\rightarrow \mathcal{A}_{*}$, describing the entanglements $\pi
_{*}=\Phi _{*}\left( \varrho _0\right) $ of the states $\varsigma =\pi
\left( I\right) $ to $\varrho =\pi _{*}\left( I\right) $ for each initial
state $\varrho _0\in \mathcal{A}_{*}^{\circ }$ with the density operator $%
\rho _0\in \mathcal{A}^{\circ }\subseteq \mathcal{L}\left( \mathcal{H}%
_0\right) $ by $\pi _{*}\left( B\right) =\mathrm{tr}_{\mathcal{F}}\kappa
\left( B\otimes I\right) \kappa ^{\dagger }$, where $\kappa =X^{\dagger
}\left( I\otimes \rho _0\right) ^{1/2}$. It is given by an entangling
transition operator $X:\mathcal{F}\otimes \mathcal{H}\rightarrow \mathcal{G}%
\otimes \mathcal{H}_0$, which is defined by a transitional amplitude
operator $Y:\mathcal{H}_0\otimes \mathcal{F}\rightarrow \mathcal{G}\otimes 
\mathcal{H}$ up to a unitary operator $U$ in $\mathcal{F}$ as 
\begin{equation*}
\left( \zeta \otimes \eta _0\right) ^{\dagger }X\left( U\xi \otimes \eta
\right) =\left( \eta _0\otimes J\xi \right) ^{\dagger }Y^{\dagger }\left(
J\zeta \otimes \eta \right) \text{.}
\end{equation*}
The dual map $\Phi :\mathcal{A}\rightarrow \mathcal{B}_{*}\otimes \mathcal{A}%
^{\circ }$ is obviously normal and completely positive, 
\begin{equation}
\Phi \left( A\right) =X\left( I\otimes A\right) X^{\dagger }\in \mathcal{B}%
_{*}\otimes \mathcal{A}^{\circ },\;\forall A\in \mathcal{A},  \label{2.9}
\end{equation}
with $\mathrm{tr}_{\mathcal{G}}\Phi \left( I\right) =I^{\circ }$, and is
called filtering map with the output states 
\begin{equation*}
\varsigma =\mathrm{tr}_{\mathcal{H}_0}\Phi \left( I\right) \left( I\otimes
\rho _0\right)
\end{equation*}
in the theory of CP flows \cite{Bel97} over $\mathcal{A}=\mathcal{A}^{\circ
} $. The operators $Y$ normalized as $\mathrm{tr}_{\mathcal{F}}Y^{\dagger
}Y=I^{\circ }$ describe $\mathcal{A}$-valued q-compound states 
\begin{equation*}
\mathrm{E}\left( B\otimes A\right) =\mathrm{tr}_{\mathcal{F}}Y^{\dagger
}\left( B\otimes A\right) Y=\mathrm{tr}_{\mathcal{G}}\left( \tilde{B}\otimes
I\right) \Phi \left( A\right) ,
\end{equation*}
defined as the normal completely positive maps $\mathcal{B}\otimes \mathcal{A%
}\rightarrow \mathcal{A}^{\circ }$ with $\mathrm{E}\left( I\otimes I\right)
=I^{\circ }$ .

If the $\mathcal{A}$-valued compound state has the diagonal form given by
the orthogonal decomposition 
\begin{equation}
\Phi \left( A\right) =\sum_n|n\rangle \mathrm{tr}_{\mathcal{F}}\Psi \left(
n\right) ^{\dagger }A\Psi \left( n\right) \langle n|,  \label{2.10}
\end{equation}
corresponding to $Y$ $=\sum_n|n\rangle \otimes \Psi \left( n\right) $, where 
$\Psi \left( n\right) :\mathcal{H}_0\otimes \mathcal{F}\rightarrow \mathcal{H%
}$, it is achieved by the d-transitional expectations 
\begin{equation*}
\Phi _{*}\left( B,\varrho _0\right) =\sum_n\langle n|B|n\rangle \Psi \left(
n\right) \left( \rho _0\otimes I\right) \Psi \left( n\right) ^{\dagger }.
\end{equation*}
The d-transitional expectations correspond to the instruments \cite{DaL71}
of the dynamical theory of quantum measurements. The elementary filters 
\begin{equation*}
\Theta _n\left( A\right) =\frac 1{\mu \left( n\right) }\mathrm{tr}_{\mathcal{%
F}}\Psi ^{\dagger }\left( n\right) A\Psi \left( n\right) ,\quad \mu \left(
n\right) =\mathrm{tr}\Psi \left( n\right) \left( \rho _0\otimes I\right)
\Psi ^{\dagger }\left( n\right)
\end{equation*}
define posterior states $\varrho _n=\varrho _0\Theta _n$ on $\mathcal{A}$
for quantum nondemolition measurements in $\mathcal{B}$, which are called
indirect if the corresponding density operators $\rho _n$ are
non-orthogonal. They describe the posterior states with orthogonal 
\begin{equation*}
\rho _n=\Psi _n\left( \rho _0\otimes I\right) \Psi _n^{\dagger },\quad \Psi
_n=\Psi \left( n\right) /\mu \left( n\right) ^{1/2}
\end{equation*}
for all $\rho _0$ iff $\Psi \left( n\right) ^{\dagger }\Psi \left( n\right)
=\delta _n^mM\left( n\right) $.

\section{Quantum Entropy via Entanglements}

As it was shown in the previous section, the diagonal entanglements describe
the classical-quantum encodings $\varkappa :\mathcal{B}\rightarrow \mathcal{A%
}_{*}$, i.e. correspondences of classical symbols to quantum, in general not
orthogonal and pure, states. As we have seen in contrast to the classical
case, not every entanglement can be achieved in this way. The general
entangled states $\varpi $ are described by the density operators $\omega
=\upsilon \upsilon ^{\dagger }$ of the form (\ref{1.9}) which are not
necessarily block-diagonal in the eigen-representation of the density
operator $\sigma $, and they cannot be achieved even by a more general
c-entanglement (\ref{2.1}). Such nonseparable entangled states are called in 
\cite{Ohy89} the quasicompound (q-compound) states, so we can call also the
quantum nonseparable correspondences the quasi-encodings (q-encodings) in
contrast to the d-correspondences, described by the diagonal entanglements.

As we shall prove in this section, the most informative for a quantum system 
$\left( \mathcal{A},\varrho \right) $ is the standard entanglement $\pi
_{*}^{\circ }=\pi _0$ of the probe system $\left( \mathcal{B}^{\circ
},\varsigma _0\right) =\left( \mathcal{A},\varrho \right) $, described in (%
\ref{1.5}). The other extreme cases of the self-dual input entanglements 
\begin{equation*}
\pi _{*}\left( A\right) =\sum_n\rho \left( n\right) ^{1/2}A\rho \left(
n\right) ^{1/2}=\pi \left( A\right) ,
\end{equation*}
are the pure c-entanglements, given by the decompositions $\rho =\sum \rho
\left( n\right) $ into pure states $\rho \left( n\right) =\eta _n\eta
_n^{\dagger }\mu \left( n\right) $. We shall see that these c-entanglements,
corresponding to the separable states 
\begin{equation}
\omega =\sum_n\eta _n\eta _n^{\dagger }\otimes \eta _n\eta _n^{\dagger }\mu
\left( n\right) ,  \label{4.0}
\end{equation}
are in general less informative then the pure d-entanglements, given in an
orthonormal basis $\left\{ \eta _n^{\circ }\right\} \subset \mathcal{H}$ by 
\begin{equation*}
\pi ^{\circ }\left( A\right) =\sum_n\eta _n^{\circ }\eta _n^{\dagger }A\eta
_n\eta _n^{\circ \dagger }\mu \left( n\right) \neq \pi _{*}^{\circ }\left(
A\right) .
\end{equation*}

Now, let us consider the entangled mutual information and quantum entropies
of states by means of the above three types of compound states. To define
the quantum mutual entropy, we need the relative entropy \cite{Lin73, Ara76}
of the compound state $\varpi $ with respect to a reference state $\varphi $
on the algebra $\mathcal{A}\otimes \mathcal{B}$. In our discrete case of the
decomposable algebras it is defined by the density operators $\omega ,\phi
\in \mathcal{B}\otimes \mathcal{A}$ of these states as 
\begin{equation}
S\left( \varpi ,\varphi \right) =\mathrm{tr}\omega \left( \ln \omega -\ln
\phi \right) .  \label{4.1}
\end{equation}
It has a positive value $S\left( \varpi ,\varphi \right) \in [0,\infty ]$ if
the states are equally normalized, say (as usually) $\mathrm{tr}\omega =1=%
\mathrm{tr}\phi $, and it can be finite only if the state $\varpi $ is
absolutely continuous with respect to the reference state $\varphi $, i.e.
iff $\varpi \left( E\right) =0$ for the maximal null-orthoprojector $E\phi
=0 $.

The mutual information $I_{\mathcal{A},\mathcal{B}}\left( \varpi \right) $
of a compound state $\varpi $ achieved by an entanglement $\pi _{*}:$ $%
\mathcal{B}\rightarrow \mathcal{A}_{*}$ with the marginals 
\begin{equation*}
\varsigma \left( B\right) =\varpi \left( B\otimes I\right) =\mathrm{tr}_{%
\mathcal{G}}B\sigma ,\;\varrho \left( A\right) =\varpi \left( I\otimes
A\right) =\mathrm{tr}_{\mathcal{H}}A\rho
\end{equation*}
is defined as the relative entropy (\ref{4.1}) with respect to the product
state $\varphi =\varsigma \otimes \varrho $: 
\begin{equation}
I_{\mathcal{A},\mathcal{B}}\left( \varpi \right) =\mathrm{tr}\omega \left(
\ln \omega -\ln \left( \sigma \otimes I\right) -\ln \left( I\otimes \rho
\right) \right) .  \label{4.3}
\end{equation}
Here the operator $\omega $ is uniquely defined by the entanglement $\pi
_{*} $ as its density in (\ref{1.8}), or the $\mathcal{G}$-transposed to the
operator $\tilde{\omega}$ in 
\begin{equation*}
\pi \left( A\right) =\kappa ^{\dagger }\left( I\otimes A\right) \kappa =%
\mathrm{tr}_{\mathcal{H}}A\tilde{\omega}.
\end{equation*}
This quantity describes an information gain in a quantum system $\left( 
\mathcal{A},\varrho \right) $ via an entanglement $\pi _{*}$ of another
system $\left( \mathcal{B},\varsigma \right) .$ It is naturally treated as a
measure of the strength of an entanglement, having zero value only for
completely disentangled states, corresponding to $\varpi =\varsigma \otimes
\varrho $.

\bigskip

\noindent
{\bf Proposition 4.1.} 
{\it
Let $\pi _{*}^{\circ }:\mathcal{B}^{\circ }\rightarrow \mathcal{A}_{*}$ be
an entanglement $\pi _{*}^{\circ }$ of a state $\varsigma _0=\pi ^{\circ
}\left( I\right) $ on a discrete decomposable algebra $\mathcal{B}^{\circ
}\subseteq \mathcal{L}\left( \mathcal{G}_0\right) $ to the state $\varrho
=\pi _{*}^{\circ }\left( I\right) $ on $\mathcal{A}$, and $\pi _{*}=\pi
_{*}^{\circ }\mathrm{K}$ be an entanglement defined as the composition with
a normal completely positive unital map $\mathrm{K}:\mathcal{B}\rightarrow 
\mathcal{B}^{\circ }$. Then $I_{\mathcal{A},\mathcal{B}}\left( \varpi
\right) \leq I_{\mathcal{A},\mathcal{B}^{\circ }}\left( \varpi _0\right) $,
where $\varpi ,\varpi _0$ are the compound states achieved by $\pi
_{*}^{\circ }$ , $\pi _{*}$ respectively. In particular, for any
c-entanglement $\pi _{*}$ to $\left( \mathcal{A},\varsigma \right) $ there
exists a not less informative d-entanglement $\pi _{*}^{\circ }=\varkappa $
with an Abelian $\mathcal{B}^{\circ }$, and the standard entanglement $\pi
_0\left( A\right) =\rho ^{1/2}A\rho ^{1/2}$ of $\varsigma _0=\varrho $ on $%
\mathcal{B}^{\circ }=\mathcal{A}$ is the maximal one in this sense.
}

\bigskip

\proof %
The first follows from the monotonicity property \cite{Lin73} 
\begin{equation}
\varpi =\mathrm{K}_{*}\varpi _0,\varphi =\mathrm{K}_{*}\varphi _0\Rightarrow
S\left( \varpi ,\varphi \right) \leq S\left( \varpi _0,\varphi _0\right)
\label{4.7}
\end{equation}
of the general relative entropy on a von Neuman algebra $\mathcal{M}$ with
respect to the predual $\mathrm{K}_{*}$ to any normal completely positive
unital map $\mathrm{K}:\mathcal{M}\rightarrow \mathcal{M}^{\circ }$. It
should be applied to the ampliation $\mathrm{K}\left( B\otimes A\right) =%
\mathrm{K}\left( B\right) \otimes A$ of the CP map $\mathrm{K}$ from $%
\mathcal{B}\rightarrow \mathcal{B}^{\circ }$ to $\mathcal{B}\otimes \mathcal{%
A}\rightarrow \mathcal{B}^{\circ }\otimes \mathcal{A}$, with the compound
state $\mathrm{K}_{*}\varpi _0=\varpi _0\left( \mathrm{K}\otimes \mathrm{I}%
\right) $ ($\mathrm{I}$ denotes the identity map $\mathcal{A}\rightarrow 
\mathcal{A}$) corresponding to the entanglement $\pi _{*}=\pi _{*}^{\circ }%
\mathrm{K}$ and $\mathrm{K}_{*}\varphi _0=\varsigma \otimes \varrho $ , $%
\varsigma =\varsigma _0\mathrm{K}$ corresponding to $\varphi _0=\varsigma
_0\otimes \varrho $.

This monotonicity property proves in particular that for any separable
compound state on $\mathcal{B}\otimes \mathcal{A}$, which is prepared by a
c-entanglement (\ref{2.1}), there exists a diagonal entanglement $\pi
_{*}^{\circ }$ to the system $\left( \mathcal{A},\varrho \right) $with the
same, or even bigger information gain (\ref{4.3}). One can take even a
classical system $\left( \mathcal{B}^{\circ },\varsigma _0\right) $, say the
diagonal sublagebra $\mathcal{B}^{\circ }$ on $\mathcal{G}_0=$ $l^2\left( 
\mathbf{N}\right) $ with the state $\varsigma _0$, induced by the measure $%
\nu $, and consider the classical-quantum correspondence (encoding) 
\begin{equation*}
\pi _{*}^{\circ }\left( B^{\circ }\right) =\sum_n\beta \left( n\right) \rho
_n\nu \left( n\right) ,\quad B^{\circ }=\sum_n|n\rangle \beta \left(
n\right) \langle n|,
\end{equation*}
prescribing the states $\varrho _n\left( A\right) =\mathrm{tr}A\rho _n$ to
the letters $n$ with the probabilities $\nu \left( n\right) $. The
information gain 
\begin{equation*}
I_{\mathcal{A},\mathcal{B}^{\circ }}\left( \varpi _0\right) =\sum_n\mu
\left( n\right) \mathrm{tr}\rho _n\left( \ln \rho _n-\ln \rho \right) .
\end{equation*}
is equal or bigger then $I_{\mathcal{A},\mathcal{B}}\left( \varpi \right) $
corresponding to $\omega =\sum_n\sigma _n\otimes \rho _n\nu \left( n\right) $
because the entanglement (\ref{2.1}) is represented as the composition $\pi
_{*}^{\circ }\mathrm{K}$ with the CP map 
\begin{equation*}
\mathrm{K}\left( B\right) =\sum_n|n\rangle \varsigma _n\left( B\right)
\langle n|,\quad B\in \mathcal{B}
\end{equation*}
into the diagonal algebra $\mathcal{B}^{\circ }$.

The inequality (\ref{4.7}) can be also applied to the standard entanglement,
corresponding to the compound state (\ref{1.6}) on $\mathcal{A}\otimes 
\mathcal{A}=\oplus _{i,k}\mathcal{A}\left( i\right) \otimes \mathcal{A}%
\left( k\right) $, where $\mathcal{A}\left( i\right) =\mathcal{L}\left( 
\mathcal{H}_i\right) $. It is described by the density operator 
\begin{equation}
\omega _0=\oplus _{i,k}\mathrm{P}_{\mathcal{A}\otimes \mathcal{A}}\left(
i,k\right) =\oplus _i\vartheta _0^i\vartheta _0^{i\dagger }\nu \left(
i\right) \text{,}  \label{4.6}
\end{equation}
with $\mathrm{P}_{\mathcal{A}\otimes \mathcal{A}}\left( i,k\right) =0$, $%
i\neq k$ concentrated on the diagonal $\oplus _i\mathcal{A}\left( i\right)
\otimes \mathcal{A}\left( i\right) $ of $\mathcal{A}\otimes \mathcal{A}$.
The amplitudes $\vartheta _0^i\in \mathcal{H}_i\otimes \mathcal{H}_i$ are
defined in (\ref{1.6}) by orthogonal components $\kappa _0\left( i\right)
=\rho \left( i\right) ^{1/2}$ of the central decomposition $\kappa
_0=\sum_n|n\rangle \otimes \kappa _0\left( n\right) $ for the standard
entangling operator $\kappa _0:\mathcal{H}\rightarrow l^2(\mathbf{N})\otimes 
\mathcal{H}$. Indeed, any entanglement $\pi _{*}\left( B\right) =\mathrm{tr}%
_{\mathcal{F}}\kappa B\kappa ^{\dagger }$ as a normal CP map $\mathcal{B}%
\rightarrow \mathcal{A}$ normalized to the density operator $\rho =\mathrm{tr%
}_{\mathcal{F}}\kappa \kappa ^{\dagger }$ can be represented as the
composition $\pi _{*}^{\circ }\mathrm{K}$ of the standard entanglement $\pi
_{*}^{\circ }=\pi _0$ on $\left( \mathcal{B}^{\circ },\varsigma _0\right)
=\left( \mathcal{A},\varrho \right) $ and a normal unital CP map $\mathrm{K}:%
\mathcal{B}\rightarrow \mathcal{A}$. The CP map $\mathrm{K}$ is defined by $%
\rho ^{1/2}\mathrm{K}\left( B\right) \rho ^{1/2}=\pi _{*}\left( B\right) $
as 
\begin{equation*}
\mathrm{K}\left( B\right) =\mathrm{tr}_{\mathcal{F}_{-}}X^{\dagger }BX,%
\mathrm{\quad }B\in \mathcal{B}
\end{equation*}
where $X$ is an operator $\mathcal{F}_{-}\otimes \mathcal{H}\rightarrow 
\mathcal{G}$, $\mathrm{tr}_{\mathcal{F}_{-}}X^{\dagger }X=I$ such that $%
\kappa =\left( I^{-}\otimes \kappa _0\right) X^{\dagger }$ is an entangling
operator for $\pi $. Thus the standard entanglement $\pi _{*}^{\circ }$
corresponds to the maximal mutual information.%
\endproof %

Note that any extreme d-entanglement 
\begin{equation*}
\pi _{*}^{\circ }\left( B\right) =\sum_n\langle n|B|n\rangle \rho _n^{\circ
}\mu \left( n\right) ,\;B\in \mathcal{B}^{\circ },
\end{equation*}
with $\rho =\sum_n\rho _n^{\circ }\mu \left( n\right) $ decomposed into pure
normalized states $\rho _n^{\circ }=\eta _n\eta _n^{\dagger }$, is maximal
among all c-entanglements in the sense $I_{\mathcal{A},\mathcal{B}}\left(
\varpi _0\right) \geq I_{\mathcal{A},\mathcal{B}}\left( \varpi \right) $.
This is because $\mathrm{tr}\rho _n^{\circ }\ln \rho _n^{\circ }=0$, and
therefore the information gain 
\begin{equation*}
I_{\mathcal{A},\mathcal{B}}\left( \varpi \right) =\sum_n\mu \left( n\right) 
\mathrm{tr}\rho _n\left( \ln \rho _n-\ln \rho \right) .
\end{equation*}
with a fixed $\pi _{*}\left( I\right) =\rho $ achieves its supremum $-%
\mathrm{tr}_{\mathcal{H}}\rho \ln \rho $ at any such extreme d-entanglement $%
\pi _{*}^{\circ }$. Thus the supremum of the information gain (\ref{4.3})
over all c-entanglements to the system $\left( \mathcal{A},\varrho \right) $
is the von Neumann entropy 
\begin{equation}
S_{\mathcal{A}}\left( \varrho \right) =-\mathrm{tr}_{\mathcal{H}}\rho \ln
\rho .  \label{4.5}
\end{equation}
It is achieved on any extreme $\pi _{*}^{\circ }$, for example given by the
maximal Abelian subalgebra $\mathcal{B}^{\circ }\subseteq \mathcal{A}$, with
the measure $\mu =\lambda $, corresponding to a Schatten decomposition $\rho
=\sum_n\eta _n^{\circ }\eta _n^{\circ \dagger }\lambda \left( n\right) $, $%
\eta _m^{\circ \dagger }\eta _n^{\circ }=\delta _n^m$. The maximal value $%
\ln \,\mathrm{rank}\mathcal{A}$ of the von Neumann entropy is defined by the
dimensionality $\mathrm{rank}\mathcal{A}=\dim \mathcal{B}^{\circ }$ of the
maximal Abelian subalgebra of the decomposable algebra $\mathcal{A}$, i.e.
by $\dim \mathcal{H}$.

\bigskip

\noindent
{\bf Definition 4.1.} 
{\it
The maximal mutual information 
\begin{equation}
H_{\mathcal{A}}\left( \varrho \right) =\sup_{\pi _{*}\left( I\right) =\rho
}I_{\mathcal{A},\mathcal{B}}\left( \varpi \right) =I_{\mathcal{A},\mathcal{B}%
^{\circ }}\left( \varpi _0\right) ,  \label{4.8}
\end{equation}
achieved on $\mathcal{B}^{\circ }=\mathcal{A}$ by the standard
q-entanglement $\pi _{*}^{\circ }\left( A\right) =\rho ^{1/2}A\rho ^{1/2}$
for a fixed state $\varrho \left( A\right) =\mathrm{tr}_{\mathcal{H}}A\rho $%
, is called q-entropy of the state $\varrho $. The differences 
\begin{equation*}
H_{\mathcal{B}|\mathcal{A}}\left( \varpi \right) =H_{\mathcal{B}}\left(
\varsigma \right) -I_{\mathcal{A},\mathcal{B}}\left( \varpi \right)
\end{equation*}
\begin{equation*}
S_{\mathcal{B}|\mathcal{A}}\left( \varpi \right) =S_{\mathcal{B}}\left(
\varsigma \right) -I_{\mathcal{A},\mathcal{B}}\left( \varpi \right)
\end{equation*}
are respectively called the q-conditional entropy on $\mathcal{B}$ with
respect to $\mathcal{A}$ and the degree of disentanglement for the compound
state $\varpi $.
}

\bigskip

Obviously, $H_{\mathcal{B}|\mathcal{A}}\left( \varpi \right) $ is positive
in contrast to the disentanglement $S_{\mathcal{B}|\mathcal{A}}\left( \varpi
\right) $, having the positive maximal value $S_{\mathcal{B}|\mathcal{A}%
}\left( \varpi \right) =S_{\mathcal{B}}\left( \varsigma \right) $ in the
case $\varpi =\varsigma \otimes \varrho $ of complete disentanglement, but
which can achieve also a negative value 
\begin{equation}
\inf_{\pi _{*}\left( I\right) =\rho }S_{\mathcal{B}|\mathcal{A}}\left(
\varpi \right) =S_{\mathcal{A}}\left( \varsigma \right) -H_{\mathcal{A}%
}\left( \varrho \right) =\sum_i\nu \left( i\right) \mathrm{tr}_{\mathcal{H}%
_i}\rho _i\ln \rho _i  \label{4.9}
\end{equation}
for the entangled states as the following theorem states. Here $\rho _i\in 
\mathcal{L}\left( \mathcal{H}_i\right) $ are the density operators of
normalized factor-states $\varrho _i=\nu \left( i\right) ^{-1}\varrho |%
\mathcal{L}\left( \mathcal{H}_i\right) $with $\nu \left( i\right) =\varrho
\left( I^i\right) $, where $I^i$ are the orthoprojectors onto $\mathcal{H}_i$%
. Obviously $H_{\mathcal{A}}\left( \varrho \right) =S_{\mathcal{A}}\left(
\varrho \right) $ if the algebra $\mathcal{A}$ is completely decomposable,
i.e. Abelian, and the maximal value $\ln \,\mathrm{rank}\mathcal{A}$ of $S_{%
\mathcal{A}}\left( \varrho \right) $ can be written as $\ln \dim \mathcal{A}$
in this case. The disentanglement $S_{\mathcal{B}|\mathcal{A}}\left( \varpi
\right) $ is always positive in this case, as well as in the case of Abelian 
$\mathcal{B}$ when $H_{\mathcal{B}|\mathcal{A}}\left( \varpi \right) =S_{%
\mathcal{B}|\mathcal{A}}\left( \varpi \right) $.

\bigskip

\noindent
{\bf Theorem 4.2.} 
{\it
Let $\mathcal{A}$ be the discrete decomposable algebra on $\mathcal{H}%
=\oplus _i\mathcal{H}_i$, with a normal state given by the density operator $%
\rho =\oplus \rho \left( i\right) $, and $\mathcal{C}\subseteq \mathcal{A}$
be its center with the state $\nu =\varrho |\mathcal{C}$ induced by the
probability distribution $\nu \left( i\right) =\mathrm{tr}\rho \left(
i\right) .$ Then the q-entropy is given by the formula 
\begin{equation}
H_{\mathcal{A}}\left( \varrho \right) =\sum_i\left( \nu \left( i\right) \ln
\nu \left( i\right) -2\mathrm{tr}_{\mathcal{H}_i}\rho \left( i\right) \ln
\rho \left( i\right) \right) ,  \label{4.10}
\end{equation}
i.e. $H_{\mathcal{A}}\left( \varrho \right) =H_{\mathcal{A}|\mathcal{C}%
}\left( \varrho \right) +H_{\mathcal{C}}\left( \nu \right) $, where $H_{%
\mathcal{C}}\left( \nu \right) =-\sum_i\nu \left( i\right) \ln \nu \left(
i\right) =S_{\mathcal{C}}\left( \nu \right) $, and 
\begin{equation*}
H_{\mathcal{A}|\mathcal{C}}\left( \varrho \right) =-2\sum_i\nu \left(
i\right) \mathrm{tr}_{\mathcal{H}_i}\rho _i\ln \rho _i=2S_{\mathcal{A}|%
\mathcal{C}}\left( \varrho \right) ,
\end{equation*}
with $\rho _i=\rho \left( i\right) /\nu \left( i\right) $. It is positive, $%
H_{\mathcal{A}}\left( \varrho \right) \in [0,\infty ]$, and if $\mathcal{A}$
is finite dimensional, it is bounded, with the maximal value $H_{\mathcal{A}%
}\left( \varrho ^{\circ }\right) =\ln \dim \mathcal{A}$ which is achieved on
the tracial $\rho ^{\circ }=\oplus \rho _i^{\circ }\nu ^{\circ }\left(
i\right) $, 
\begin{equation*}
\rho _i^{\circ }=\left( \dim \mathcal{H}_i\right) ^{-1}I^i,\quad \nu ^{\circ
}\left( i\right) =\dim \mathcal{A}\left( i\right) /\dim \mathcal{A},
\end{equation*}
where $\dim \mathcal{A}\left( i\right) =\left( \dim \mathcal{H}_i\right) ^2$%
, $\dim \mathcal{A}=\sum_i\dim \mathcal{A}\left( i\right) $.
}

\bigskip

\proof %
The q-entropy $H_{\mathcal{A}}\left( \varrho \right) $ is the supremum (\ref
{4.8}) of the mutual information (\ref{4.3}) which is achieved on the
standard entanglement, corresponding to the density operator (\ref{4.6}) of
the standard compound state (\ref{1.6}) with $\mathcal{B}=\mathcal{A}$, $%
\sigma =\rho $. Thus $H_{\mathcal{A}}\left( \rho \right) =I_{\mathcal{A},%
\mathcal{A}}\left( \varpi _0\right) $, where 
\begin{eqnarray*}
I_{\mathcal{A},\mathcal{A}}\left( \varpi _0\right) &=&\mathrm{tr}_{\mathcal{H%
}\otimes \mathcal{H}}\omega _0\left( \ln \omega _0-\ln \left( \rho \otimes
I\right) -\ln \left( I\otimes \rho \right) \right) \\
&=&\sum_i\nu \left( i\right) \ln \nu \left( i\right) -2\mathrm{tr}\rho \ln
\rho =-\sum_i\nu \left( i\right) \left( \ln \nu \left( i\right) +2\mathrm{tr}%
_{\mathcal{H}_i}\rho _i\ln \rho _i\right) .
\end{eqnarray*}
Here we used that $\mathrm{tr}\omega _0\ln \omega _0=\sum_i\nu \left(
i\right) \ln \nu \left( i\right) $ due to 
\begin{equation*}
\omega _0\ln \omega _0=\oplus _{i,k}\mathrm{P}_{\mathcal{A}\otimes \mathcal{A%
}}\left( i,k\right) \ln \mathrm{P}_{\mathcal{A}\otimes \mathcal{A}}\left(
i,k\right) =\oplus _i\nu \left( i\right) \vartheta _0^i\vartheta
_0^{i\dagger }\ln \nu \left( i\right)
\end{equation*}
for the orthogonal diagonal decomposition (\ref{4.6}) of $\omega _0$ into
one-dimensional orthoprojectors $\vartheta _0^i\vartheta _0^{i\dagger }=%
\mathrm{P}_{\mathcal{A}\otimes \mathcal{A}}\left( i,i\right) /\nu \left(
i\right) $, and that $\mathrm{tr}\rho \ln \rho =\sum_i\nu \left( i\right)
\left( \ln \nu \left( i\right) -S_{\mathcal{A}_i}\left( \varrho _i\right)
\right) $ due to 
\begin{equation*}
\rho \ln \rho =\oplus _i\mathrm{P}_{\mathcal{A}}\left( i\right) \ln \mathrm{P%
}_{\mathcal{A}}\left( i\right) =\oplus _i\nu \left( i\right) \rho _i\left(
\ln \nu \left( i\right) +\ln \rho _i\right)
\end{equation*}
for the orthogonal decomposition $\rho =\oplus _i\nu \left( i\right) \mathrm{%
P}_{\mathcal{A}\left( i\right) }$, where $\mathrm{P}_{\mathcal{A}\left(
i\right) }=\mathrm{P}_{\mathcal{A}}\left( i\right) /\nu \left( i\right)
=\rho _i$, $\nu \left( i\right) =\mathrm{trP}_{\mathcal{A}}\left( i\right) $%
, $\mathrm{P}_{\mathcal{A}}\left( i\right) =\sum_k\mathrm{tr}_{\mathcal{H}}%
\mathrm{P}_{\mathcal{A}\otimes \mathcal{A}}\left( i,k\right) =\rho \left(
i\right) $.

Thus $H_{\mathcal{A}}\left( \varrho \right) =H_{\mathcal{A}|\mathcal{C}%
}\left( \varrho \right) +H_{\mathcal{C}}\left( \nu \right) =2S_{\mathcal{A}|%
\mathcal{C}}\left( \varrho \right) +S_{\mathcal{C}}\left( \nu \right) $ is
positive, and it is bounded by 
\begin{eqnarray*}
C_{\mathcal{A}} &=&\sup_\nu \sum_i\nu \left( i\right) \left( 2\sup_{\varrho
_i}S_{\mathcal{A}\left( i\right) }\left( \varrho _i\right) -\ln \nu \left(
i\right) \right) \\
&=&-\inf_\nu \sum_i\nu \left( i\right) \left( \ln \nu \left( i\right) -2\ln 
\mathrm{\dim }\mathcal{H}_i\right) =\ln \dim \mathcal{A}.
\end{eqnarray*}
Here we used the fact that the supremum of von Neumann entropies 
\begin{equation*}
S_{\mathcal{A}\left( i\right) }\left( \varrho _i\right) =-\sum_i\mathrm{tr}_{%
\mathcal{H}_i}\rho _i\ln \rho _i
\end{equation*}
for the simple algebras $\mathcal{A}\left( i\right) =\mathcal{L}\left( 
\mathcal{H}_i\right) $ with $\dim \mathcal{A}\left( i\right) =\left( \dim 
\mathcal{H}_i\right) ^2<\infty $ is achieved on the tracial density
operators $\rho _i=\left( \dim \mathcal{H}_i\right) ^{-1}I^i\equiv \rho
_i^{\circ }$, and the infimum of the relative entropy 
\begin{equation*}
S\left( \nu ,\nu ^{\circ }\right) =\sum_i\nu \left( i\right) \left( \ln \nu
\left( i\right) -\ln \nu ^{\circ }\left( i\right) \right) ,
\end{equation*}
where $\nu ^{\circ }\left( i\right) =\dim \mathcal{A}\left( i\right) /\dim 
\mathcal{A}$, is zero, achieved at $\nu =\nu ^{\circ }$.%
\endproof %

\section{Quantum Channel and its Q-Capacity}

Let $\mathcal{H}_0$ be a Hilbert space describing a quantum input system and 
$\mathcal{H}$ describe its output Hilbert space. A quantum channel is an
affine operation sending each input state defined on $\mathcal{H}_0$ to an
output state defined on $\mathcal{H}$ such that the mixtures of states are
preserved. A deterministic quantum channel is given by a linear isometry $Y%
\mathrm{:\mathcal{H}}_0\rightarrow \mathrm{\mathcal{H}}$ with $Y^{\dagger
}Y=I^{\circ }$ ($I^{\circ }$ is the identify operator in $\mathrm{\mathcal{H}%
}_0$) such that each input state vector $\eta \in \mathrm{\mathcal{H}}_0$, $%
\left\| \eta \right\| =1$ is transmitted into an output state vector $Y\eta
\in \mathcal{H}$, $\left\| Y\eta \right\| =1$. The orthogonal mixtures $\rho
_0=\sum_n\mu \left( n\right) \rho _n^{\circ }$ of the pure input states $%
\rho _n^{\circ }=\eta _n^{\circ }\eta _n^{\circ \dagger }$ are sent into the
orthogonal mixtures $\rho =\sum_n\mu \left( n\right) \rho _n$ of the
corresponding pure states $\rho _n=Y\rho _n^{\circ }Y^{\dagger }$.

A noisy quantum channel sends pure input states $\varrho _0$ into mixed ones 
$\varrho =\Lambda ^{*}\left( \varrho _0\right) $ given by the dual $\Lambda
^{*}$ to a normal completely positive unital map $\Lambda :\mathcal{A}%
\rightarrow \mathcal{A}_0$, 
\begin{equation*}
\Lambda \left( A\right) =\mathrm{tr}_{\mathcal{F}_1}Y^{\dagger }AY,\mathrm{%
\quad }A\in \mathrm{\mathcal{A}}
\end{equation*}
where $Y$ is a linear operator from $\mathrm{\mathcal{H}}_0\otimes \mathcal{F%
}_{+}$ to $\mathrm{\mathcal{H}}$ with $\mathrm{tr}_{\mathcal{F}%
_{+}}Y^{\dagger }Y=I^{\circ }$, and $\mathcal{F}_{+}$ is a separable Hilbert
space of quantum noise in the channel. Each input mixed state $\varrho _0$
on $\mathcal{A}^{\circ }\subseteq \mathcal{L}\left( \mathcal{H}_0\right) $
is transmitted into an output state $\varrho =\varrho _0\Lambda $ given by
the density operator 
\begin{equation*}
\Lambda _{*}\left( \rho _0\right) =Y\left( \rho _0\otimes I^{+}\right)
Y^{\dagger }\in \mathcal{A}_{*}
\end{equation*}
for each density operator $\rho _0\in \mathcal{A}_{*}^{\circ }$, where $%
I^{+} $ is the identity operator in $\mathcal{F}_{+}$. Without loss of
generality we can assume that the input algebra $\mathcal{A}^{\circ }$ is
the smallest decomposable algebra, generated by the range $\Lambda \left( 
\mathcal{A}\right) $ of the given map $\Lambda $.

The input entanglements $\varkappa :\mathcal{B}\rightarrow \mathcal{A}%
_{*}^{\circ }$ described as normal CP maps with $\varkappa \left( I\right)
=\varrho _0$, define the quantum correspondences (q-encodings) of probe
systems $\left( \mathcal{B},\varsigma \right) $, $\varsigma =\varkappa
^{*}\left( I\right) $, to $\left( \mathcal{A}^{\circ },\varrho _0\right) $.
As it was proven in the previous section, the most informative is the
standard entanglement $\varkappa =\pi _{*}^{\circ }$, at least in the case
of the trivial channel $\Lambda =\mathrm{I}$. This extreme input
q-entanglement 
\begin{equation*}
\pi ^{\circ }\left( A^{\circ }\right) =\rho _0^{1/2}A^{\circ }\rho
_0^{1/2}=\pi _{*}^{\circ }\left( A^{\circ }\right) ,\quad A^{\circ }\in 
\mathcal{A}^{\circ },
\end{equation*}
corresponding to the choice $\left( \mathcal{B},\varsigma \right) =\left( 
\mathcal{A}^{\circ },\varrho _0\right) $, defines the following density
operator 
\begin{equation}
\omega =\left( \mathrm{I}\otimes \Lambda \right) _{*}\left( \omega _q^{\circ
}\right) ,\quad \omega _q^{\circ }=\bigoplus_i\left( \vartheta _0^\iota
\vartheta _0^{\iota \dagger }\right) \nu _0\left( i\right)  \label{3.1}
\end{equation}
of the input-output compound state $\varpi _q^{\circ }\Lambda $ on $\mathcal{%
A}^{\circ }\otimes \mathcal{A}$. It is given by the central decomposition $%
\rho _0=\oplus \rho _{0i}\nu _0\left( i\right) $ of the density operator $%
\rho _0\in \mathcal{A}_{*}^{\circ }=\oplus \mathcal{T}\left( \mathcal{H}%
_{0i}\right) $, with the amplitudes $\vartheta _0^i\in \mathcal{H}%
_{0i}^{\otimes 2}$ defined by $\tilde{\vartheta}_0^\iota =\rho _{0i}^{1/2}$.
The other extreme cases of the self-dual input entanglements, the pure
c-entanglements corresponding to (\ref{4.0}), can be less informative then
the d-entanglements, given by the decompositions $\rho _0=\sum \rho _0\left(
n\right) $ into pure states $\rho _0\left( n\right) =\eta _n\eta _n^{\dagger
}\mu \left( n\right) $. They define the density operators 
\begin{equation}
\omega =\left( \mathrm{I}\otimes \Lambda \right) _{*}\left( \omega _d^{\circ
}\right) ,\quad \omega _d^{\circ }=\sum_n\eta _n^{\circ }\eta _n^{\circ
\dagger }\otimes \eta _n\eta _n^{\dagger }\mu _0\left( n\right) ,
\label{3.2}
\end{equation}
of the $\mathcal{A}^{\circ }\otimes \mathcal{A}$-compound state $\varpi
_d^{\circ }\Lambda $, which are known as the Ohya compound states $\varpi
_o^{\circ }\Lambda $ \cite{Ohy83} in the case 
\begin{equation*}
\rho _0\left( n\right) =\eta _n^{\circ }\eta _n^{\circ \dagger }\lambda
_0\left( n\right) ,\quad \eta _m^{\circ \dagger }\eta _n^{\circ }=\delta
_n^m,
\end{equation*}
of orthogonality of the density operators $\rho _0\left( n\right) $
normalized to the eigen-values $\lambda _0\left( n\right) $ of $\rho _0$.
They are described by the input-output density operators 
\begin{equation}
\omega =\left( \mathrm{I}\otimes \Lambda \right) _{*}\left( \omega _o^{\circ
}\right) ,\quad \omega _o^{\circ }=\sum_n\eta _n^{\circ }\eta _n^{\circ
\dagger }\otimes \eta _n^{\circ }\eta _n^{\circ \dagger }\lambda _0\left(
n\right) ,  \label{3.3}
\end{equation}
coinciding with (\ref{3.1}) in the case of Abelian $\mathcal{A}^{\circ }$.
These input-output compound states $\varpi $ are achieved by compositions $%
\lambda =\pi ^{\circ }\Lambda $, describing the entanglements $\lambda ^{*}$
of the extreme probe system $\left( \mathcal{B}^{\circ },\varsigma _0\right)
=\left( \mathcal{A}^{\circ },\varrho _0\right) $ to the output $\left( 
\mathcal{A},\varrho \right) $ of the channel.

If $\mathrm{K}:\mathcal{B}\rightarrow \mathcal{B}^{\circ }$ is a normal
completely positive unital map 
\begin{equation*}
\mathrm{K}\left( B\right) =\mathrm{tr}_{\mathcal{F}_{-}}X^{\dagger }BX,\quad
B\in \mathcal{B},
\end{equation*}
where $X$ is a bounded operator $\mathcal{F}_{-}\otimes \mathcal{G}%
_0\rightarrow \mathcal{G}$ with $\mathrm{tr}_{\mathcal{F}_{-}}X^{\dagger
}X=I^{\circ }$, the compositions $\varkappa =\pi _{*}^{\circ }\mathrm{K}$, $%
\pi _{*}=\Lambda _{*}\varkappa $ are the entanglements of the probe system $%
\left( \mathcal{B},\varsigma \right) $ to the channel input $\left( \mathcal{%
A}^{\circ },\varrho _0\right) $ and to the output $\left( \mathcal{A}%
,\varrho \right) $ via this channel. The state $\varsigma =\varsigma _0%
\mathrm{K}$ is given by 
\begin{equation*}
\mathrm{K}_{*}\left( \sigma _0\right) =X\left( I^{-}\otimes \sigma _0\right)
X^{\dagger }\in \mathcal{B}_{*}
\end{equation*}
for each density operator $\sigma _0\in \mathcal{B}_{*}^{\circ }$, where $%
I^{-}$ is the identity operator in $\mathcal{F}_{-}$. The resulting
entanglement $\pi _{*}=\lambda _{*}\mathrm{K}$ defines the compound state $%
\varpi =\varpi _0\left( \mathrm{K}\otimes \Lambda \right) $ on $\mathcal{B}%
\otimes \mathcal{A}$ with 
\begin{equation*}
\varpi _0\left( B^{\circ }\otimes A^{\circ }\right) =\mathrm{tr}\tilde{B}%
^{\circ }\pi ^{\circ }\left( A^{\circ }\right) =\mathrm{tr}\upsilon
_0^{\dagger }\left( B^{\circ }\otimes A^{\circ }\right) \upsilon _0.
\end{equation*}
on $\mathcal{B}^{\circ }\otimes \mathcal{A}^{\circ }$. Here $\upsilon _0:%
\mathcal{F}_0\rightarrow \mathcal{G}_0\otimes \mathrm{\mathcal{H}}_0$ is the
amplitude operator, uniquely defined by the input compound state $\varpi
_0\in \mathcal{B}_{*}^{\circ }\otimes \mathcal{A}_{*}^{\circ }$ up to a
unitary operator $U^{\circ }$ on $\mathcal{F}_0$, and the effect of the
input entanglement $\varkappa $ and the output channel $\Lambda $ can be
written in terms of the amplitude operator of the state $\varpi $ as 
\begin{equation*}
\upsilon =\left( X\otimes Y\right) \left( I^{-}\otimes \upsilon _0\otimes
I^{+}\right) U
\end{equation*}
up to a unitary operator $U$ in $\mathcal{F}=\mathcal{F}_{-}\otimes \mathcal{%
F}_0\otimes \mathcal{F}_{+}$. Thus the density operator $\omega =\upsilon
\upsilon ^{\dagger }$ of the input-output compound state $\varpi $ is given
by $\varpi _0\left( \mathrm{K}\otimes \Lambda \right) $ with the density 
\begin{equation}
\left( \mathrm{K}\otimes \Lambda \right) _{*}\left( \omega _0\right) =\left(
X\otimes Y\right) \omega _0\left( X\otimes Y\right) ^{\dagger },  \label{3.4}
\end{equation}
where $\omega _0=\upsilon _0\upsilon _0^{\dagger }$.

Let $\mathcal{K}_q$ be the convex set of normal completely positive maps $%
\varkappa :\mathcal{B}\rightarrow \mathcal{A}_{*}^{\circ }$ normalized as $%
\mathrm{tr}\varkappa \left( I\right) =1$, and $\mathcal{K}_q^{\circ }$ be
the convex subset $\left\{ \varkappa \in \mathcal{K}_q:\varkappa \left(
I\right) =\varrho _0\right\} $. Each $\varkappa \in \mathcal{K}_q^{\circ }$
can be decomposed as $\pi _{*}^{\circ }\mathrm{K}$, where $\pi _{*}^{\circ
}=\pi ^{\circ }$ is the standard entanglement on $\left( \mathcal{A}^{\circ
},\varrho _0\right) $, and $\mathrm{K}$ is a normal unital CP map $\mathcal{B%
}\rightarrow \mathcal{A}^{\circ }$. Further let $\mathcal{K}_c$ be the
convex set of the maps $\varkappa $, dual to the input maps of the form (\ref
{2.1}), described by the combinations 
\begin{equation}
\varkappa \left( B\right) =\sum_n\varsigma \left( B\right) \rho _0\left(
n\right) .  \label{3.6}
\end{equation}
of the primitive maps $\varkappa _n:B\mapsto \varsigma _n\left( B\right)
\rho _0\left( n\right) $, and $\mathcal{K}_d$ be the subset of the diagonal
decompositions 
\begin{equation}
\varkappa \left( B\right) =\sum_n\langle n|B|n\rangle \rho _0\left( n\right)
.  \label{3.5}
\end{equation}
As in the first case $\mathcal{K}_c^{\circ }$ and $\mathcal{K}_d^{\circ }$
denote the convex subsets corresponding to a fixed $\varkappa \left(
I\right) =\varrho _0$, and each $\varkappa \in \mathcal{K}_c^{\circ }$ can
be represented as $\pi _{*}^{\circ }\mathrm{K}$, where $\pi _{*}^{\circ }$
is a d-entanglement, which can be always be made pure by a proper choice of
the CP map $\mathrm{K}:\mathcal{B}\rightarrow \mathcal{A}^{\circ }$.
Furthermore let $\mathcal{K}_o$ ($\mathcal{K}_o^{\circ }$) be the subset of
all decompositions (\ref{3.6}) with orthogonal $\rho _0\left( n\right) $
(and fixed $\sum_n\rho _0\left( n\right) =\rho _0$): 
\begin{equation*}
\quad \rho _0\left( m\right) \rho _0\left( n\right) =0,\,m\neq n.
\end{equation*}
Each $\varkappa \in \mathcal{K}_o^{\circ }$ can be also represented as $\pi
_{*}^{\circ }\mathrm{K}$, where $\pi _{*}^{\circ }$ is a diagonal pure
o-entanglement $\mathcal{B}\rightarrow \mathcal{A}^{\circ }$.

Now, let us maximize the entangled mutual information for a given quantum
channel $\Lambda $ and a fixed input state $\varrho _0$ by means of the
above four types of compound states. The mutual information (\ref{4.3}) was
defined in the previous section by the density operators of the compound
state $\varpi $ on $\mathcal{B}\otimes \mathcal{A}$, and the product-state $%
\varphi =\varsigma \otimes \varrho $ of the marginals $\varsigma ,\varrho $
for $\varpi $. In each case 
\begin{equation*}
\varpi =\varpi _0\left( \mathrm{K}\otimes \Lambda \right) ,\quad \varphi
=\varphi _0\left( \mathrm{K}\otimes \Lambda \right) ,
\end{equation*}
where $\mathrm{K}$ is a CP map $\mathcal{B}\rightarrow \mathcal{B}^{\circ }$%
, $\varpi _0$ is one of the corresponding extreme compound states $\varpi
_q^{\circ }$, $\varpi _c^{\circ }=\varpi _d^{\circ }$, $\varpi _o^{\circ }$
on $\mathcal{A}^{\circ }\otimes \mathcal{A}^{\circ }$, and $\varphi
_0=\varrho _0\otimes \varrho _0$. The density operator $\omega =\left( 
\mathrm{K}\otimes \Lambda \right) _{*}\left( \omega _0\right) $ is written
in (\ref{3.4}), and $\phi =\sigma \otimes \rho $ can be written as 
\begin{equation*}
\phi =\varkappa _{*}\left( I\right) \otimes \lambda _{*}\left( I\right) ,
\end{equation*}
where $\lambda _{*}=\Lambda _{*}\pi _{*}^{\circ }$.

\bigskip

\noindent
{\bf Proposition 5.1.} 
{\it
The entangled mutual informations achieve the following maximal values 
\begin{equation}
\sup_{\varkappa \in \mathcal{K}_q^{\circ }}I_{\mathcal{A},\mathcal{B}}\left(
\varpi \right) =I_q\left( \varrho _0,\Lambda \right) :=I_{\mathcal{A},%
\mathcal{A}^{\circ }}\left( \varpi _q^{\circ }\Lambda \right) ,  \label{3.7}
\end{equation}
\begin{equation*}
I_c\left( \varrho _0,\Lambda \right) =\sup_{\varkappa \in \mathcal{K}%
_c^{\circ }}I_{\mathcal{A},\mathcal{B}}\left( \varpi \right) =\sup_{\varpi
_d^{\circ }}I_{\mathcal{A},\mathcal{A}^{\circ }}\left( \varpi _d^{\circ
}\Lambda \right) =I_d\left( \varrho _0,\Lambda \right) ,
\end{equation*}
\begin{equation}
\sup_{\varkappa \in \mathcal{K}_o^{\circ }}I_{\mathcal{A},\mathcal{B}}\left(
\varpi \right) =I_o\left( \varrho _0,\Lambda \right) :=\sup_{\varpi
_o^{\circ }}I_{\mathcal{A},\mathcal{A}^{\circ }}\left( \varpi _o^{\circ
}\Lambda \right) ,  \label{3.8}
\end{equation}
where $\varpi _{\bullet }^{\circ }$ are the corresponding extremal input
entangled states on $\mathcal{A}^{\circ }\otimes \mathcal{A}^{\circ }$ with
marginals $\varrho _0$. They are ordered as 
\begin{equation}
I_q\left( \varrho _0,\Lambda \right) \geq I_c\left( \varrho _0,\Lambda
\right) =I_d\left( \varrho _0,\Lambda \right) \geq I_o\left( \varrho
_0,\Lambda \right) .  \label{3.9}
\end{equation}
}

\bigskip

\proof %
Due to the monotonicity 
\begin{equation*}
I_{\mathcal{A},\mathcal{B}}\left( \varpi _d^{\circ }\left( \mathrm{K}\otimes
\Lambda \right) \right) \leq I_{\mathcal{A},\mathcal{A}^{\circ }}\left(
\varpi _d^{\circ }\left( \mathrm{I}\otimes \Lambda \right) \right)
\end{equation*}
the supremum over all c-entanglements $\varkappa \in \mathcal{K}_c^{\circ }$
coinsides with the supremum over $\mathcal{K}_d^{\circ }\subset \mathcal{K}%
_c^{\circ }$ which is achieved on the pure d-entanglements on $\left( 
\mathcal{A}^{\circ },\varrho _0\right) $ corresponding to the extreme
compound states $\varpi _d^{\circ }$. By the same monotonicity arguments we
can get the equalities (\ref{3.7}) and (\ref{3.8}). The entanglements $%
\varkappa \in \mathcal{K}_q^{\circ }$ can be written as 
\begin{equation*}
\varkappa \left( B\right) =\sum_{m,n}\langle m|B|n\rangle \chi \left(
m\right) \chi \left( n\right) ^{\dagger }
\end{equation*}
in a basis $\left\{ |n\rangle \right\} \subset \mathcal{G}$ for the Schatten
decompositions $\sigma =\sum_n|n\rangle \mu \left( n\right) \langle n|$
corresponding to weakly orthogonal amplitude operators $\chi \left( n\right)
=\left( \langle n|X\otimes I\right) \left( I^{-}\otimes \upsilon _0\right) :$%
\textrm{\ } 
\begin{equation*}
\mathrm{tr}\chi \left( m\right) \chi \left( n\right) ^{\dagger }=\mu \left(
n\right) \delta _n^m.
\end{equation*}
The maps $\varkappa \in \mathcal{K}_d^{\circ }$ can be written as 
\begin{equation*}
\varkappa \left( B\right) =\sum_n\langle n|B|n\rangle \chi \left( n\right)
\chi \left( n\right) ^{\dagger }
\end{equation*}
corresponding to stronger orthogonal amplitude operators 
\begin{equation*}
\chi \left( m\right) \chi \left( n\right) ^{\dagger }=\rho _0\left( n\right)
\delta _n^m,
\end{equation*}
defining not necessarily orthogonal decompositions $\rho _0=\sum_n\rho
_0\left( n\right) $. The extreme maps $\varkappa \in \mathcal{K}_o^{\circ }$
can be written as 
\begin{equation*}
\varkappa \left( B\right) =\sum_n\langle n|B|n\rangle \chi \left( n\right)
\chi \left( n\right) ^{\dagger }
\end{equation*}
with amplitude operators $\chi \left( n\right) $, satisfying the second
orthogonality condition 
\begin{equation*}
\chi \left( n\right) ^{\dagger }\chi \left( m\right) =\mu \left( n\right)
\tau _n^{\circ }\delta _n^m,
\end{equation*}
where $\tau _n^{\circ }$ are density operators in $\mathcal{F}_0$ with the
traces $\mathrm{tr}\tau _n^{\circ }=1$. Thus, the inequalities in (\ref{3.9}%
) follow from $\mathcal{K}_q\supseteq \mathcal{K}_c\supseteq \mathcal{K}%
_d\supseteq \mathcal{K}_o$.%
\endproof %

We shall denote the maximal informations $I_c\left( \varrho _0,\Lambda
\right) =I_d\left( \varrho _0,\Lambda \right) $ simply as $I\left( \varrho
_0,\Lambda \right) $.

\bigskip

\noindent
{\bf Definition 5.1.} 
{\it
The supremums 
\begin{equation*}
C_q\left( \Lambda \right) =\sup_{\varkappa \in \mathcal{K}_q}I_{\mathcal{A},%
\mathcal{B}}\left( \varpi \right) =\sup_{\varrho _0}I_q\left( \varrho
_0,\Lambda \right) ,\;
\end{equation*}
\begin{equation}
\sup_{\varkappa \in \mathcal{K}_c}I_{\mathcal{A},\mathcal{B}}\left( \varpi
\right) =C\left( \Lambda \right) :=\sup_{\varrho _0}I\left( \varrho
_0,\Lambda \right) ,\;  \label{3.10}
\end{equation}
\begin{equation*}
C_o\left( \Lambda \right) =\sup_{\varkappa \in \mathcal{K}_o}I_{\mathcal{A},%
\mathcal{B}}\left( \varpi \right) =\sup_{\varrho _0}I_o\left( \varrho
_0,\Lambda \right) ,\;
\end{equation*}
are called the q-, c- or d-, and o-capacities respectively for the quantum
channel defined by a normal unital CP map $\Lambda :\mathcal{A}\rightarrow 
\mathcal{A}^{\circ }$.
}

\bigskip

Obviously the capacities (\ref{3.10}) satisfy the inequalities 
\begin{equation*}
C_o\left( \Lambda \right) \leq C\left( \Lambda \right) \leq C_q\left(
\Lambda \right) .
\end{equation*}

\bigskip

\noindent
{\bf Theorem 5.2.} 
{\it
Let $\Lambda \left( A\right) =Y^{\dagger }AY$ be a unital CP map $\mathcal{A}%
\rightarrow \mathcal{A}^{\circ }$ describing a quantum deterministic
channel. Then 
\begin{equation*}
I\left( \varrho _0,\Lambda \right) =I_o\left( \varrho _0,\Lambda \right)
=S\left( \varrho _0\right) ,\quad I_q\left( \varrho _0,\Lambda \right)
=S_q\left( \varrho _0\right) ,
\end{equation*}
where $S_q\left( \varrho _0\right) =H_{\mathcal{A}^{\circ }}\left( \varrho
_0\right) $, and thus in this case 
\begin{equation*}
C\left( \Lambda \right) =C_o\left( \Lambda \right) =\ln \mathrm{rank}%
\mathcal{A}^{\circ },\quad C_q\left( \Lambda \right) =\ln \dim \mathcal{A}%
^{\circ }.
\end{equation*}
}

\bigskip

\proof %
It was proved in the previous section for the case of the identity channel $%
\Lambda =\mathrm{I}$, and thus it is also valied for any isomorphism $%
\Lambda $ described by a unitary operator $Y$. In the case of non-unitary $Y$
we can use the identity 
\begin{equation*}
\mathrm{tr}Y\left( \rho _0\otimes I^{+}\right) Y^{\dagger }\ln Y\left( \rho
_0\otimes I^{+}\right) Y^{\dagger }=\mathrm{tr}R\left( \omega _0\otimes
I^{+}\right) \ln R\left( \omega _0\otimes I^{+}\right) ,
\end{equation*}
where $R=Y^{\dagger }Y$. Due to this $S\left( \varrho _0\Lambda \right) =-%
\mathrm{tr}R\left( \rho _0\otimes I^{+}\right) \ln R\left( \rho _0\otimes
I^{+}\right) $, and $S\left( \varpi _0\left( \mathrm{K}\otimes \Lambda
\right) \right) =$ 
\begin{equation*}
-\mathrm{tr}\left( S\otimes R\right) \left( I^{-}\otimes \omega _0\otimes
I^{+}\right) \ln \left( S\otimes R\right) \left( I^{-}\otimes \omega
_0\otimes I^{+}\right) ,
\end{equation*}
where $S=X^{\dagger }X$. Thus $S\left( \varrho _0\Lambda \right) =S\left(
\varrho _0\right) $, $S\left( \varpi _0\left( \mathrm{K}\otimes \Lambda
\right) \right) =S\left( \varpi _0\left( \mathrm{K}\otimes \mathrm{I}\right)
\right) $ if $Y^{\dagger }Y=I$, and 
\begin{eqnarray*}
I_{\mathcal{A},\mathcal{B}}\left( \varpi _0\left( \mathrm{K}\otimes \Lambda
\right) \right) &=&S\left( \varsigma _0\mathrm{K}\right) +S\left( \varrho
_0\right) -S\left( \varpi _0\left( \mathrm{K}\otimes \mathrm{I}\right)
\right) \\
&\leq &S\left( \varsigma _0\right) +S\left( \varrho _0\right) -S\left(
\varpi _0\right) =I_{\mathcal{A}^{\circ },\mathcal{B}^{\circ }}\left( \varpi
_0\right)
\end{eqnarray*}
for any normal unital CP map $\mathrm{K}:\mathcal{B}\rightarrow \mathcal{B}%
^{\circ }$ and a compound state $\varpi _0$ on $\mathcal{B}^{\circ }\otimes 
\mathcal{A}^{\circ }$. The supremum (\ref{3.7}), which is achieved at the
standard entanglement, corresponding to $\varpi _0=\varpi _q$, coincides
with q-entropy $H_{\mathcal{A}^{\circ }}\left( \varrho _0\right) $, and the
supremum (\ref{3.8}), coinciding with $S_{\mathcal{A}^{\circ }}\left(
\varrho _0\right) $, is achieved for a pure o-entanglement, corresponding to 
$\varpi _0=\varpi _o$ given by any Schatten decomposition for $\rho _0$.
Moreover, the entropy $H_{\mathcal{A}^{\circ }}\left( \varrho _0\right) $ is
also achieved by any pure d-entanglement, corresponding to $\varpi _0=\varpi
_d$ given by any extreme decomposition for $\rho _0$, and thus is the
maximal mutual information $I\left( \varrho _0,\Lambda \right) $ in the case
of deterministic $\Lambda $. Thus the capacity $C\left( \Lambda \right) $ of
the deterministic channel is given by the maximum $C_o=\ln \dim \mathcal{H}%
_0 $ of the von Neumann entropy $S_{\mathcal{A}^{\circ }}$, and the
q-capacity $C_q\left( \Lambda \right) $ is equal $C_{\mathcal{A}^{\circ
}}=\ln \dim \mathcal{A}^{\circ }$.%
\endproof %

In the general case d-entanglements can be more informative than
o-entanglements as it can be shown on an example of a quantum noisy channel
for which 
\begin{equation*}
I\left( \varrho _0,\Lambda \right) >I_o\left( \varrho _0,\Lambda \right)
,\quad C\left( \Lambda \right) >C_o\left( \Lambda \right) .
\end{equation*}
The last equalities of the above theorem will be related to the work on
entropy by Voiculescu \cite{Voi}.

\end{document}